\documentclass[aoas]{imsart}

\RequirePackage{amsthm,amsmath,amsfonts,amssymb}
\RequirePackage[authoryear]{natbib}
\RequirePackage[colorlinks,citecolor=blue,urlcolor=blue]{hyperref}
\RequirePackage{graphicx, multirow}
\RequirePackage{placeins}

\graphicspath{{Figure/}{./}}

\startlocaldefs
\theoremstyle{plain}

\theoremstyle{definition}

\endlocaldefs

\begin{document}

\begin{frontmatter}
\title{Multivariate Continuous-Time Autoregressive Moving Average Processes for Astronomical Multiband Time Series}
\runtitle{MCARMA for Astronomical Multiband Time Series}

\begin{aug}
%%%%%%%%%%%%%%%%%%%%%%%%%%%%%%%%%%%%%%%%%%%%%%%
%% Only one address is permitted per author. %%
%% Only division, organization and e-mail is %%
%% included in the address.                  %%
%% Additional information such as            %%
%% identifying the corresponding author must %%
%% be included in in the Acknowledgments     %%
%% section if necessary.                     %%
%% ORCID can be inserted by command:         %%
%% \orcid{0000-0000-0000-0000}               %%
%%%%%%%%%%%%%%%%%%%%%%%%%%%%%%%%%%%%%%%%%%%%%%%
\author[A]{\fnms{Izak}~\snm{Schmidlkofer} \ead[label=e1]{schmidiz@oregonstate.edu}}
\author[A]{\fnms{Zhirui}~\snm{Hu}\ead[label=e2]{zhirui.hu@oregonstate.edu}}
\author[B]{\fnms{Lishan}~\snm{Shi}\ead[label=e3]{lfs5712@psu.edu}}
\author[C]{\fnms{Weixiang}~\snm{Yu}\ead[label=e4]{wuy10@psu.edu}}
\author[B, C, D, E]{\fnms{Hyungsuk}~\snm{Tak}\ead[label=e5]{tak@psu.edu}\orcid{0000-0000-0000-0000}}

%%%%%%%%%%%%%%%%%%%%%%%%%%%%%%%%%%%%%%%%%%%%%%
%% Addresses                                %%
%%%%%%%%%%%%%%%%%%%%%%%%%%%%%%%%%%%%%%%%%%%%%%
%\address[A]{???\printead[presep={,\ }]{e1}}

%\address[B]{???\printead[presep={,\ }]{e2,e3}}
%\end{aug}

\address[A]{Department of Statistics,
Oregon State University \printead[presep={ ,\ }]{e1,e2}}

\address[B]{Department of Statistics,
Pennsylvania State University \printead[presep={,\ }]{e3,e5}}

\address[C]{Department of Astronomy and Astrophysics,
Pennsylvania State University \printead[presep={,\ }]{e4}}

\address[D]{Institute for Computational and Data Sciences,
Pennsylvania State University}

\address[E]{The Research Institute of Basic Sciences,
Seoul National University}

\end{aug}

\begin{abstract}
Large-scale astronomical surveys provide unprecedented volumes of
multivariate time-series observations obtained through multiple
optical filters. We develop a structured multivariate continuous-time
autoregressive moving average (MCARMA) framework for multi-band time
series with irregular sampling, heteroscedastic measurement errors,
and partially observed bands. The framework allows band-specific
stochastic dynamics while modeling cross-band dependence through
correlated Brownian driving processes, with state-space and spectral
representations enabling likelihood-based inference and interpretation
of the fitted stochastic dynamics. We develop a two-stage estimation
procedure in which a numerically stabilized preliminary fit initializes
subsequent maximum likelihood estimation. Simulations show that
higher-order stochastic structure can be recovered when its
characteristic features are adequately resolved, but can become weakly
identifiable because of limited temporal resolution or near
pole--zero cancellation. Joint multivariate estimation improves
parameter recovery in 23 of 27 settings and spectral recovery in 26
of 27 settings relative to separate single-band fits. Three Sloan
Digital Sky Survey Stripe~82 quasars, respectively favoring
MCARMA$(1,0)$, MCARMA$(2,0)$, and MCARMA$(2,1)$, illustrate how
joint multiband modeling uses cross-band dependence to inform marginal dynamics and can yield different model-order and spectral inference. The
methodology is implemented in the Python package \texttt{mcarma}.

\end{abstract}


\end{frontmatter}
%%%%%%%%%%%%%%%%%%%%%%%%%%%%%%%%%%%%%%%%%%%%%%
%% Please use \tableofcontents for articles %%
%% with 50 pages and more                   %%
%%%%%%%%%%%%%%%%%%%%%%%%%%%%%%%%%%%%%%%%%%%%%%
%\tableofcontents

\section{Introduction}
\label{sec1}

The increasing scale of time-domain astronomical surveys has created
a growing need for statistical methods that can jointly model multiple
dependent time series observed irregularly and with measurement
uncertainty. Gaussian processes provide a natural framework for such
data because they accommodate irregular observation times,
heteroscedastic measurement errors, and partially observed bands
\citep{rybicki1992time1,2023ARA&A..61..329A}. In astronomy, one of
the most widely used models in this class is the damped random walk
(DRW) process \citep{kelly2009variations}, also known as the
Ornstein--Uhlenbeck process or a continuous-time autoregressive
process of order one. Its stochastic parameters have been related to
astrophysical properties such as black hole mass and luminosity
\citep{kelly2009variations,macleod2010modeling,kozlowski2010quantifying},
and the DRW has been used extensively to characterize quasar
variability and as a stochastic model in time-delay inference
\citep[see][for a recent review]{2024ApJS..275...30T}.

The first-order dynamics of the DRW can nevertheless be restrictive.
Departures from DRW behavior have been reported on short timescales
\citep{2011ApJ...743L..12M,2013ApJ...765..106Z}, and a single
characteristic timescale may not capture the diversity of quasar
variability
\citep{2014MNRAS.439..703G,2015MNRAS.451.4328K,2023ApJ...950...37M}.
Continuous-time autoregressive moving average (CARMA) processes
provide a flexible extension, allowing multiple characteristic
timescales and power spectral densities with richer frequency-domain
structure
\citep{kelly2014flexible,2019PASP..131f3001M,2022ApJ...936..132Y,
2025ApJ...992..130Y}.

Most astronomical applications, however, fit CARMA models separately
to individual photometric bands. Joint modeling is attractive because
multi-band observations of the same source are dependent, allowing
information to be shared when observations in individual bands are
sparse or noisy and enabling direct inference on cross-band
dependence. \citet{2020AJ....160..265H} applied a structured
multivariate DRW formulation to astronomical multi-band data, with
band-specific first-order dynamics coupled through correlated
stochastic drivers and likelihood-based inference accommodating
irregular sampling, heteroscedastic measurement errors, and partially
observed bands. Extending this structured formulation from first-order
DRW dynamics to higher-order CARMA dynamics provides a natural
framework for modeling richer multi-band stochastic variability.

Multivariate CARMA (MCARMA) processes have an established theoretical
and statistical literature
\citep{marquardt2007mCARMA,schlemm2012multivariate,schlemm2012qMLE,
BROCKWELL2013217,fasen2017information}. Particularly relevant here,
\citet{tomasson2018} considered stationary Wiener-driven Gaussian
MCARMA processes in state-space form and used Kalman filtering for
likelihood inference from irregular and nonsynchronous multivariate
observations. Thus, Wiener-driven Gaussian MCARMA processes,
state-space likelihood evaluation, and accommodation of irregular and
nonsynchronous observations are not themselves new. Our contribution
is instead to develop and study a structured MCARMA methodology
tailored to astronomical multi-band time series with irregular
sampling, heteroscedastic measurement errors, and partially observed
bands, with particular emphasis on joint estimation of band-specific
stochastic dynamics and their frequency-domain representation.

We consider a structured subclass of Wiener-driven Gaussian MCARMA
processes with diagonal autoregressive and moving-average coefficient
matrices, allowing each band to have its own temporal dynamics while
inducing cross-band dependence through correlated Brownian drivers.
This structure separates marginal dynamics from cross-band dependence
and yields a matrix-valued spectral density characterizing marginal
power spectral densities and cross-band spectral dependence. The model
is formulated in state-space form so that its Gaussian likelihood can
be evaluated by Kalman filtering with computational cost linear in the
number of unique observation times for fixed model dimension. For
inference, we develop a root-based parameterization that enforces
stationarity and minimum-phase constraints and facilitates
interpretation of characteristic dynamical scales and near pole--zero
cancellation. We further develop a two-stage estimation procedure in
which a numerically stabilized and regularized preliminary fit
initializes subsequent maximization of the unregularized
log-likelihood. Model order is selected using the finite-sample
corrected Akaike information criterion
\citep{akaike1974new,10.1093/biomet/76.2.297}, and the methodology is
implemented in the publicly available Python package \texttt{mcarma}
\citep{mcarma2026}.

Our simulation study investigates when higher-order MCARMA dynamics
can be recovered, when their parameters become weakly identified, and
how much is gained by fitting dependent bands jointly rather than
separately. The simulations retain key features of astronomical
observations, including irregular sampling, seasonal gaps,
heteroscedastic measurement errors, and partially observed bands,
while systematically varying stochastic dynamics and
measurement-noise levels. They reveal two distinct sources of weak
identifiability: observational limitations when characteristic
timescales fall outside the temporal range resolved by the data, and
structural limitations when an autoregressive pole and a moving-average
zero nearly cancel. Joint multivariate estimation improves parameter
recovery in 23 of 27 settings and spectral recovery in 26 of 27
settings relative to separate single-band fits. Importantly, accurate
spectral recovery can persist even when individual dynamical
parameters are weakly identified, making the power spectral density a
useful process-level and scientifically interpretable summary of the
fitted stochastic process.

We illustrate these findings using three quasars from the Sloan
Digital Sky Survey Stripe~82 sample \citep{macleod2012},
selected to represent MCARMA$(1,0)$, MCARMA$(2,0)$, and
MCARMA$(2,1)$ fits. Because MCARMA$(1,0)$ is the multivariate DRW,
the first quasar has fitted marginal DRW spectra under the joint
model. The second exhibits the steeper high-frequency decline of a
second-order autoregressive spectrum, whereas in the third the
moving-average component modifies the spectral shape and restores the
$f^{-2}$ asymptotic decline. These examples therefore illustrate
three distinct forms of marginal stochastic behavior within the
structured MCARMA family.

We also compare the joint fits with the band-by-band approach commonly
used in astronomical applications, selecting a univariate CARMA order
separately by AICc for each band. For the first two quasars, the
independently selected single-band fits yield substantially more
heterogeneous spectral behavior across bands than the joint fits,
whereas the two approaches agree more closely for the third. These
comparisons illustrate an important consequence of multiband modeling:
when an individual band is only weakly informative, joint MCARMA
estimation can use cross-band dependence to inform its marginal
stochastic dynamics. The examples additionally illustrate how
observational resolution affects the interpretation of fitted
timescales and provide inference on cross-band dependence unavailable
from separate single-band analyses. Together, the simulations and
case studies show when higher-order multivariate stochastic dynamics
can be recovered and how joint versus separate analysis can affect
their inferred spectral structure under realistic astronomical
observing conditions.

The remainder of the paper is organized as follows.
Section~\ref{sec2} introduces the structured MCARMA family, and
Section~\ref{sec3} develops its state-space representation and
Kalman-filter likelihood. Section~\ref{sec4} presents the root-based
representation, parameter constraints, and identifiability
considerations. Section~\ref{sec:estimation} describes estimation,
uncertainty quantification, and model selection.
Section~\ref{sec:simulation} presents the simulation study,
Section~\ref{sec:sdss} presents the SDSS case studies, and
Section~\ref{sec:conclusion} concludes.

\section{Structured MCARMA Processes}
\label{sec2}

We develop a structured subclass of Wiener-driven Gaussian multivariate continuous-time autoregressive moving average (MCARMA) processes in which individual components may exhibit distinct temporal dynamics while sharing stochastic variation across components. The formulation builds on the MCARMA framework of \citet{tomasson2018}, with additional structure tailored to joint modeling of component-specific dynamics and cross-component dependence. Rather than allowing cross-component dependence to enter through unrestricted matrix-valued autoregressive (AR) and moving-average (MA) operators, we use component-specific dynamic operators and induce cross-component dependence through correlated Brownian drivers. In the astronomical setting considered here, the components correspond to photometric bands, so this structure separates band-specific stochastic dynamics from cross-band dependence.

\subsection{Model formulation and structured dependence}
\label{sec2:model}

Let
\[
\mathbf{X}(t)=\{X_1(t),\ldots,X_k(t)\}^{\top}
\]
denote a $k$-dimensional zero-mean continuous-time stochastic process.
A Wiener-driven Gaussian MCARMA$(p,q)$ process, with $p>q\geq0$, can be represented
formally as
\begin{equation}
A(D)\mathbf{X}(t)=M(D)\,d\mathbf{B}(t),
\label{eq:mcarma}
\end{equation}
where $D=d/dt$ and
\[
A(z)=A_0+A_1z+\cdots+A_{p-1}z^{p-1}+I_kz^p,
\qquad
M(z)=I_k+M_1z+\cdots+M_qz^q
\]
are $k\times k$ AR and MA matrix polynomials. Here $I_k$ is the
$k\times k$ identity matrix and $\mathbf{B}(t)$ is a $k$-dimensional
Brownian motion satisfying
\[
\operatorname{Cov}\{d\mathbf{B}(t)\}=V\,dt
\]
for a positive-semidefinite $k\times k$ matrix $V$.
Equation~\eqref{eq:mcarma} is defined rigorously through its
state-space representation below.

In a general Wiener-driven Gaussian MCARMA model, the matrices in $A(z)$ and $M(z)$
need not be diagonal, so cross-component dependence can enter through
the dynamic operators as well as through $V$. We instead restrict
\[
A_i=\operatorname{diag}(\alpha_{i,1},\ldots,\alpha_{i,k}),
\qquad
M_l=\operatorname{diag}(\beta_{l,1},\ldots,\beta_{l,k}),
\]
for $i=0,\ldots,p-1$ and $l=1,\ldots,q$. For component $j$,
\begin{align*}
A_j(z)
&=z^p+\alpha_{p-1,j}z^{p-1}+\cdots+\alpha_{1,j}z+\alpha_{0,j},
\\
M_j(z)
&=\beta_{q,j}z^q+\cdots+\beta_{1,j}z+1.
\notag
\end{align*}
Thus each component has its own marginal CARMA$(p,q)$ dynamics,
whereas dependence among components is induced through the correlated
Brownian drivers. The multivariate DRW of
\citet{2020AJ....160..265H} is recovered when $(p,q)=(1,0)$.

For interpretation, write
\begin{equation*}
V=D_\sigma R D_\sigma,
\qquad
D_\sigma=\operatorname{diag}(\sigma_1,\ldots,\sigma_k),
\end{equation*}
where $R=(\rho_{jl})$ is a correlation matrix, so that
$V_{jl}=\sigma_j\sigma_l\rho_{jl}$. The diagonal elements of $V$
determine stochastic driving amplitudes and its off-diagonal elements
determine dependence between the drivers. For numerical optimization
we instead parameterize $V$ through a Cholesky factor, as described in
Section~\ref{sec4}.

\subsection{State-space and covariance representations}
\label{sec2:statespace}

The structured MCARMA process has a finite-dimensional linear
state-space representation
\begin{equation}
d\mathbf{Z}(t)=F\mathbf{Z}(t)\,dt+G\,d\mathbf{B}(t),
\qquad
\mathbf{X}(t)=H\mathbf{Z}(t),
\label{eq:ss}
\end{equation}
where $\mathbf{Z}(t)$ is a \(kp\)-dimensional latent state vector, and the matrices $F$, $G$, and $H$ are determined by the AR and MA
coefficient matrices; their explicit forms are given in Appendix~\ref{sec:supp.statespace}. A stationary solution exists when all
eigenvalues of $F$ have negative real parts, which under the structured
model is equivalent to requiring all roots of each $A_j(z)$ to lie in
the open left half-plane.

Let $P_\infty=\operatorname{Cov}\{\mathbf{Z}(t)\}$ denote the
stationary state covariance. It satisfies the continuous-time
Lyapunov equation
\begin{equation*}
FP_\infty+P_\infty F^\top+GVG^\top=0.
\end{equation*}
Consequently, for $h\geq0$,
\begin{equation*}
\operatorname{Cov}\{\mathbf{X}(t+h),\mathbf{X}(t)\}
=
He^{Fh}P_\infty H^\top.
\end{equation*}
Because the driving process is Gaussian and the state-space system is linear, the resulting process also admits an equivalent covariance-based multi-output Gaussian process representation; see, for example, \citet{das2026modeling} for this perspective in astronomical multi-band time-series modeling. Direct evaluation of the resulting
dense covariance likelihood is possible, but the state-space
representation permits sequential likelihood evaluation by Kalman
filtering. Further covariance identities and the explicit
multi-output Gaussian process representation are given in Appendix~\ref{sec:supp.statespace}.

\subsection{Spectral representation}
\label{sec2:spectral}

Let
\[
\mathcal{H}(z)=A(z)^{-1}M(z)
\]
denote the transfer matrix. The matrix-valued spectral density is
\begin{equation*}
P(\omega)
=
\mathcal{H}(i\omega)V\mathcal{H}(i\omega)^{*},
\end{equation*}
where $^*$ denotes conjugate transpose. Throughout, $\omega$ denotes
angular frequency in radians per unit time. When frequencies are
reported in cycles per day, as in the numerical results, we write
\begin{equation*}
f=\frac{\omega}{2\pi}.
\end{equation*}
Under this convention the covariance and spectral density are related
by
\[
C(h)=\frac{1}{2\pi}\int_{-\infty}^{\infty}
P(\omega)e^{i\omega h}\,d\omega.
\]
For every $\omega$, $P(\omega)$ is Hermitian and positive
semidefinite. Its diagonal entries are marginal power spectral
densities (PSDs), and its off-diagonal entries are cross-spectral
densities.

Under the diagonal dynamic structure, define
\[
h_j(z)=\frac{M_j(z)}{A_j(z)}.
\]
Then
\begin{equation*}
P_{jl}(\omega)
=
h_j(i\omega)V_{jl}\overline{h_l(i\omega)},
\end{equation*}
and in particular
\begin{equation*}
P_{jj}(\omega)
=
V_{jj}\frac{|M_j(i\omega)|^2}{|A_j(i\omega)|^2}.
\end{equation*}
Thus each marginal PSD has the functional form of a univariate
CARMA$(p,q)$ spectrum, although its fitted parameters and spectrum
need not coincide with those obtained by fitting the band
independently.

A normalized measure of frequency-domain dependence is the
coherence\footnote{Following the convention commonly used in the
astronomical literature (e.g., \citealt{2014A&ARv..22...72U}), we
refer to $\gamma_{jl}^2(\omega)$ as the \emph{coherence}; it is often
called magnitude-squared or squared coherence elsewhere.}
\citep{2014A&ARv..22...72U},
\[
\gamma_{jl}^2(\omega)
=
\frac{|P_{jl}(\omega)|^2}{P_{jj}(\omega)P_{ll}(\omega)}
=
\rho_{jl}^2.
\]
Hence coherence is constant with frequency under the adopted
structured dependence, even though the cross-spectrum itself
generally varies with frequency. More general non-diagonal MCARMA
models need not have frequency-independent coherence. Cross-spectral
phase and related properties are discussed in Appendix~\ref{sec:supp.spectral}.

Finally, when the leading MA coefficient is nonzero,
\begin{equation*}
P_{jj}(\omega)
=
O\!\left(|\omega|^{-2(p-q)}\right),
\qquad |\omega|\rightarrow\infty.
\end{equation*}
Therefore, MCARMA$(1,0)$ has the familiar $-2$ high-frequency slope,
whereas MCARMA$(2,0)$ has a $-4$ asymptotic slope. MCARMA$(2,1)$
again has an asymptotic slope of $-2$, although its MA zero can
substantially modify the spectral shape over finite frequencies.
These spectral summaries are especially useful when individual
dynamic parameters are weakly identified, motivating our evaluation
of both parameter and spectral recovery in Section~\ref{sec:simulation}.

\section{Likelihood Evaluation via Kalman Filtering}
\label{sec3}

Let $t_1<\cdots<t_n$ denote the ordered union of observation times
across all components, and set $\Delta_i=t_i-t_{i-1}$. The exact
state transition implied by Equation~\eqref{eq:ss} is
\begin{equation*}
\mathbf{Z}_i
=
\Phi_i\mathbf{Z}_{i-1}+\boldsymbol{\eta}_i,
\qquad
\Phi_i=e^{F\Delta_i},
\end{equation*}
where $\boldsymbol{\eta}_i\sim N(\mathbf{0},Q_i)$. Under stationarity,
\begin{equation*}
Q_i=P_\infty-\Phi_iP_\infty\Phi_i^\top.
\end{equation*}
Thus each actual time gap is handled directly without interpolation
or temporal binning.

Let $\boldsymbol{\mu}=(\mu_1,\ldots,\mu_k)^\top$ denote the
component-specific mean levels. At time $t_i$, let
$S_i\in\{0,1\}^{k_i^\ast\times k}$ select the $k_i^\ast$ observed
components. The observation model is
\begin{equation*}
\mathbf{y}_i^\ast
=
S_i\boldsymbol{\mu}
+
S_iH\mathbf{Z}_i
+
\boldsymbol{\epsilon}_i^\ast,
\qquad
\boldsymbol{\epsilon}_i^\ast
\sim
N\!\left(
\mathbf{0},
R_i^\ast
\right),
\end{equation*}
where $R_i^\ast$ is diagonal and contains the known,
observation-specific measurement-error variances. This formulation
accommodates arbitrary partial and nonsynchronous observation
patterns without changing the latent state dimension.

Given the predicted state mean $\widehat{\mathbf{Z}}_{i|i-1}$ and
covariance $P_{i|i-1}$, define the innovation and its covariance by
\begin{align*}
\mathbf{x}_i^\ast
&=
\mathbf{y}_i^\ast-S_i\boldsymbol{\mu}
-S_iH\widehat{\mathbf{Z}}_{i|i-1},
\\
W_i^\ast
&=
S_iHP_{i|i-1}H^\top S_i^\top+R_i^\ast.
\end{align*}
The Gaussian log-likelihood is then
\begin{equation*}
\log L(\boldsymbol{\theta};\mathbf{y}^\ast)
=
-\frac{1}{2}
\sum_{i=1}^{n}
\left\{
k_i^\ast\log(2\pi)
+\log|W_i^\ast|
+(\mathbf{x}_i^\ast)^\top
(W_i^\ast)^{-1}\mathbf{x}_i^\ast
\right\}.
\end{equation*}
The complete prediction and updating recursions, including the
stationary initialization and numerically stable covariance update,
are given in Appendix~\ref{sec:supp.kalman}.

Under the structured MCARMA$(p,q)$ model, the number of free
parameters is
\begin{equation}
d=k(p+q+2)+\frac{k(k-1)}{2},
\label{eq:npar}
\end{equation}
including $k$ means, $kp$ AR parameters, $kq$ MA parameters, $k$
stochastic-amplitude parameters, and $k(k-1)/2$ cross-component
dependence parameters.

For fixed $k$, $p$, and $q$, likelihood evaluation scales linearly
with the number of unique observation times. A more detailed
complexity accounting and comparison with the equivalent dense
Gaussian-process likelihood are given in Appendix~\ref{sec:supp.kalman},
with empirical CPU-time benchmarks provided in Supplementary
Section~2.7. The same state-space representation also permits
latent-process reconstruction through the Rauch--Tung--Striebel
smoother \citep{rauch1965maximum}, with numerical illustrations
provided in Appendix~\ref{sec:supp.simulation.rts}.

\section{Parameterization and Identifiability}
\label{sec4}

Direct optimization over AR and MA polynomial coefficients is
inconvenient because stationarity and minimum phase impose nonlinear
root-location restrictions. We therefore distinguish between
\emph{optimization coordinates}, which enforce the required
constraints, and \emph{root-based quantities}, which are used for
interpretation and identifiability diagnostics.

\subsection{Root-based interpretation and constraints}
\label{sec4:roots}

For band $j$, let $\lambda_{r,j}$ denote the $r$th root of the
AR polynomial $A_j(z)$ and $z_{m,j}$ the $m$th root of the MA
polynomial $M_j(z)$. We refer to $\lambda_{r,j}$ as an AR pole and
$z_{m,j}$ as an MA zero. For an AR pole $\lambda_{r,j}$, define
\[
\kappa_{r,j}=-\operatorname{Re}(\lambda_{r,j}),
\qquad
\tau_{r,j}=\kappa_{r,j}^{-1}.
\]
Stationarity requires $\kappa_{r,j}>0$. For a real pole,
$\tau_{r,j}$ is the exponential decay timescale; for a complex
conjugate pair, the real part determines damping while the imaginary
part determines oscillation frequency. The pole modulus
$|\lambda_{r,j}|$ provides a characteristic angular-frequency scale.

For MCARMA$(1,0)$,
\[
A_j(z)=z+\tau_j^{-1},
\]
so each band has one characteristic decay timescale and the model
reduces to the multivariate DRW. For the second-order models considered
here, we write
\begin{equation*}
A_j(z)
=
z^2+2\zeta_j\omega_{n,j}z+\omega_{n,j}^2,
\end{equation*}
where $\omega_{n,j}>0$ is the natural angular frequency and
$\zeta_j>0$ is the damping ratio. The two AR poles, corresponding to
the roots of $A_j(z)$, are
\[
\lambda_{1,j}
=
\omega_{n,j}
\left(
-\zeta_j-\sqrt{\zeta_j^2-1}
\right),
\qquad
\lambda_{2,j}
=
\omega_{n,j}
\left(
-\zeta_j+\sqrt{\zeta_j^2-1}
\right).
\]
Thus $\zeta_j<1$, $\zeta_j=1$, and $\zeta_j>1$ correspond to
underdamped, critically damped, and overdamped dynamics,
respectively. Underdamping alone does not imply a nonzero-frequency
spectral peak; for MCARMA$(2,0)$ such a peak requires
$\zeta_j<1/\sqrt{2}$.

For MCARMA$(2,1)$,
\[
M_j(z)=1+b_{1,j}z,
\qquad
z_{1,j}=-b_{1,j}^{-1}.
\]
The MA zero modifies the spectral response without introducing an
additional AR timescale. We impose a minimum-phase convention by
requiring all MA zeros to lie in the open left half-plane. The
optimization parameterization expresses the AR and MA polynomials as
products of positive linear and quadratic factors and represents
those positive factor coefficients on the logarithmic scale.
Stationarity and minimum phase are therefore maintained throughout
optimization. The complete factorization and transformations are
given in Appendix~\ref{sec:supp.parameterization}.

The driving covariance is parameterized as
\[
V=LL^\top,
\]
where $L$ is lower triangular with positive diagonal elements; the
diagonal elements are represented on a logarithmic scale. This
guarantees positive definiteness for every finite optimization
parameter vector. For interpretation, the fitted $V$ is converted
back to stochastic amplitudes $\sigma_j=\sqrt{V_{jj}}$ and
correlations
$\rho_{jl}=V_{jl}/\sqrt{V_{jj}V_{ll}}$.

\subsection{Weak identifiability}
\label{sec4:identifiability}

For an AR pole $\lambda_{r,j}$ and MA zero $z_{m,j}$, define their
separation by
\[
d_{rm,j}=|\lambda_{r,j}-z_{m,j}|.
\]
If $d_{rm,j}=0$, the corresponding AR and MA factors cancel exactly,
so that the nominal MCARMA$(2,1)$ representation reduces to a
nonminimal representation of a lower-order MCARMA$(1,0)$ process.
More practically, when $d_{rm,j}$ is small but nonzero, the process
can closely resemble this lower-order process, and the likelihood can
be relatively insensitive to separate movements of the pole and zero
that preserve their near cancellation. This produces a structural
source of weak identifiability.
Structural near cancellation is distinct from weak identification
caused by the observational design. Sparse sampling, a short
baseline, or large measurement errors can leave characteristic
timescales outside the range resolved by the observations; improving
the cadence, baseline, or precision can increase information in such
cases. By contrast, exact pole--zero cancellation is intrinsically
nonminimal, and near cancellation can leave the individual pole and
zero weakly identified even under favorable sampling.

Both mechanisms matter for interpretation, and neither is specific to the MA component. Selection of a second AR pole does not imply that both characteristic timescales fall within the range of frequencies the observations resolve, and selection of MCARMA(2,1) indicates support for the additional spectral flexibility of the MA component without implying that the pole and the zero are separately well resolved. In either case, different parameter combinations can imply very similar transfer functions and power spectral densities over the frequencies supported by the data. Accordingly, the numerical studies evaluate both parameter recovery and recovery of the implied stochastic process through its power spectral density.

\section{Estimation and Model Selection}
\label{sec:estimation}

Optimizing the likelihood function for higher-order models can be sensitive to initialization when dynamical scales are weakly constrained, poles and zeros nearly cancel, or the driving covariance approaches the boundary of the positive-definite cone. We therefore
use a two-stage procedure that guarantees both numerical stabilization and
the statistical property of the final estimator.

\subsection{Two-stage maximum likelihood estimation}
\label{sec:pmle}

In the first stage we maximize
\begin{equation*}
Q(\boldsymbol{\theta})
=
\log L_{\rm reg}(\boldsymbol{\theta};\mathbf{y}^{\ast})
-
\mathcal{J}(\boldsymbol{\theta}),
\end{equation*}
where $\mathcal{J}$ contains soft penalties that discourage dynamical
scales poorly informed by the observing design, together with a soft
box on the scale of the driving covariance, and $L_{\rm reg}$ uses a
mild diagonal loading of the driving covariance for numerical
stability. Concretely, $\mathcal{J}$ keeps the moduli of the AR and MA
roots within the frequency range the observing design resolves and
places a floor on the damping of each AR pole, which prevents a pole
from sliding toward the stability boundary, where the unpenalized
first-stage objective has no interior optimum. Stationarity, minimum
phase, and positive definiteness are not penalized; they are enforced
directly by the parameterization.

Near pole--zero cancellation is not penalized. That configuration is
left to model selection. As all penalties and the covariance
loading are removed in the second stage, the reported likelihood and
the AICc comparison across orders are those of the ordinary model, so a
near-cancelling higher-order fit is judged only by whether its
additional parameters pay for themselves. The exact penalties,
design-informed reference scales, covariance loading, starting-value
construction, and optimization details are given in Supplementary
Sections~1.6--1.7.

Let $\widehat{\boldsymbol{\theta}}_{\rm init}$ denote the best
first-stage solution. In the second stage all penalties and covariance
loading are removed, and we maximize the ordinary likelihood,
\begin{equation*}
\widehat{\boldsymbol{\theta}}
=
\arg\max_{\boldsymbol{\theta}}
~\log L(\boldsymbol{\theta};\mathbf{y}^{\ast}),
\end{equation*}
initialized at $\widehat{\boldsymbol{\theta}}_{\rm init}$. Thus the
first stage is only a numerical device for obtaining a stable
initialization. Reported parameter estimates, uncertainty summaries,
and model comparisons are based on the ordinary, unpenalized
likelihood. Both stages primarily use gradient-based optimization, with derivative-free fallback procedures when needed. Starting values, the convergence criteria, and the treatment of fits that do not meet them are given in Appendix~\ref{sec:supp.optimization}.

\subsection{Uncertainty quantification}
\label{sec:uncertainty}

For a converged ordinary-likelihood fit, let
\[
\widehat{\mathcal I}
=
-\nabla^2
\log L(\boldsymbol{\theta};\mathbf{y}^{\ast})
\big|_{\boldsymbol{\theta}=\widehat{\boldsymbol{\theta}}}
\]
denote the observed information in the unconstrained optimization
coordinates. When it is positive definite and sufficiently well
conditioned,
\begin{equation*}
\widehat{\operatorname{Var}}
(\widehat{\boldsymbol{\theta}})
=
\widehat{\mathcal I}^{-1}.
\end{equation*}
Standard errors for derived quantities such as polynomial
coefficients, characteristic timescales, driving amplitudes,
correlations, and coherence are obtained by the delta method.
Near-singular curvature is itself a diagnostic of weak local
identification; in such cases Hessian-based standard errors are not
treated as reliable. Their finite-sample calibration is evaluated in
the simulations, with additional details reported in Appendix~\ref{sec:supp.uncertainty}.

\subsection{Model-order selection}
\label{sec:modelselection}

We compare three MCARMA model orders, \((1,0)\), \((2,0)\), and \((2,1)\), using
the Akaike information criterion with a finite-sample correction
(AICc; \citealt{akaike1974new, 10.1093/biomet/76.2.297}). For $d$ free parameters
and
\begin{equation*}
N=\sum_{i=1}^{n}k_i^\ast,
\end{equation*}
the total number of observed scalar band measurements, we compute
\begin{equation*}
\mathrm{AICc}
=
-2\log L(\widehat{\boldsymbol{\theta}};\mathbf{y}^{\ast})
+
2d
+
\frac{2d(d+1)}{N-d-1}.
\end{equation*}
Here $N$ counts individual observed band measurements rather than
unique observation times, an important distinction for partially
observed or asynchronous multivariate data. The parameter count $d$
is given in Equation~\eqref{eq:npar}. We select the candidate with the
smallest AICc and report $\Delta\mathrm{AICc}$ relative to the
preferred model where useful.

AICc is evaluated at the final unpenalized maximum likelihood
estimate, so the first-stage penalties and covariance loading do not
enter the criterion. Every fitted order enters the comparison. When
the second stage does not converge at a given order, the first-stage
solution is retained and rescored under the same ordinary likelihood,
and it is recorded as unconverged rather than reported as a maximum
likelihood estimate. Its log-likelihood is then a lower bound on that
order's maximum, so the criterion is conservative against such a
candidate: it can lose a comparison it should have won, but it cannot
win one it should have lost. Supplementary
Section~1.7.3 reports the convergence census for the analyses in
Section~\ref{sec:sdss} and the effect of restricting attention to
objects that converge at every fitted order. A sensitivity analysis comparing AICc with a
generalized information criterion \citep{konishi1996gic} is reported in Appendix~\ref{sec:supp.selection}; the simulations support the use of AICc for
the analyses in this paper.

\section{Simulation Studies}
\label{sec:simulation}

We use simulations of five-band quasar time series to evaluate the statistical performance of the proposed MCARMA framework under known generative models. The simulations retain key features of astronomical multi-band observations, including irregular sampling, seasonal gaps, heteroscedastic measurement errors, and partially observed bands. Holding this astronomy-motivated observing design fixed while systematically varying the underlying stochastic dynamics and measurement-noise level allows us to isolate their effects on MCARMA inference. These simulation experiments complement the real-data analysis of three SDSS quasars in Section~\ref{sec:sdss}, where the underlying stochastic processes and their parameters are unknown. Additional coefficient-wise recovery results, alternative spectral-error summaries, reconstruction experiments, uncertainty-calibration diagnostics, sensitivity analyses, and computational benchmarks are reported in Appendix~\ref{sec:supp.controlled}.

All simulation-based statements about estimation accuracy are made under known generative parameters. Model-order recovery is evaluated after fitting all three candidate orders, whereas parameter and spectral comparisons are evaluated at the generative order unless stated otherwise. This distinction separates failure to choose the correct model order from difficulty estimating an otherwise correctly specified model.

\subsection{Simulation design and evaluation criteria}
\label{sec:simulation.design}

We simulate five-band quasar time series from MCARMA$(1,0)$, MCARMA$(2,0)$, and MCARMA$(2,1)$ processes. For each second-order model, we consider underdamped, critically damped, and overdamped regimes. Cross-band dependence is generated according to
\begin{equation*}
\rho_{j\ell}=0.9^{|j-\ell|},
\end{equation*}
so that neighboring bands are strongly correlated while dependence decreases smoothly with band separation. All simulated data sets follow the same astronomy-motivated observing
design, with irregular sampling, seasonal gaps, and observations in
only one band at each epoch, reflecting the staggered multi-filter
sampling expected from surveys such as the Vera C. Rubin Observatory
Legacy Survey of Space and Time \citep{ivezi2019lsst};
observation times are drawn separately for each data set. A data set contains 2439--2674 scalar observations (median 2527), or about 500 observations per band. Measurement uncertainties are varied to produce nominal variability signal-to-noise ratios (S/N) of approximately 10, 4, and 2, defined as the stationary standard deviation of the intrinsic stochastic process divided by the per-epoch measurement uncertainty.

For each combination of generative order, damping regime, and variability S/N, we generate 33 independent data sets. MCARMA$(1,0)$ has no damping classification; its three nominal damping scenarios at a given S/N use the same 33 generative processes observed through independent realizations. Across S/N levels the generative parameters are held fixed. The complete simulation corpus contains 891 data sets. For every data set, we fit MCARMA$(1,0)$, MCARMA$(2,0)$, and MCARMA$(2,1)$ using the two-stage procedure of Section~\ref{sec:estimation} and select the preferred order using AICc computed from the final ordinary likelihood. At the generative order, we additionally fit the corresponding univariate CARMA model separately to each band.

We evaluate model-order recovery by the proportion of data sets for which AICc selects the generative order. To evaluate process-level estimation accuracy, we emphasize recovery of the marginal power spectral densities (PSDs), which summarize the combined effect of the fitted dynamical parameters. Let $P_{jj}(f)$ and $\widehat P_{jj}(f)$ denote the generative and estimated marginal PSDs. The shape-normalized spectral error (SNSE) is
\begin{equation*}
\operatorname{SNSE}_j
=
\frac{
\displaystyle\int_{f_0}^{f_1}
[\widehat{\bar P}_{jj}(f)-\bar P_{jj}(f)]^2\,df
}{
\displaystyle\int_{f_0}^{f_1}[\bar P_{jj}(f)]^2\,df
},
\end{equation*}
where $\bar P_{jj}(f)=P_{jj}(f)/P_{jj}(f_0)$ and $\widehat{\bar P}_{jj}(f)=\widehat P_{jj}(f)/\widehat P_{jj}(f_0)$. We use $f_0=10^{-3.5}$ and $f_1=10^{-0.3}$ cycles day$^{-1}$ and average SNSE over all five  bands. The normalization removes overall spectral level, so SNSE primarily measures recovery of spectral shape. Appendix~\ref{sec:supp.simulation.multivariate} reports coefficient-wise relative errors
and additional spectral-accuracy results, which support the same
qualitative conclusion about the benefit of joint fitting.

\subsection{Model-order recovery and weak identifiability}
\label{sec:simulation.order}

Table~\ref{tab:param_recovery} reports the proportion of simulated data sets for which AICc selects the generative MCARMA order. Model-order recovery depends not only on nominal model complexity and measurement noise but also strongly on the dynamical configuration represented by the generative parameters.

\begin{table}[t]
\centering
\caption{Correct-order recovery rates in the simulation study. Each second-order entry is based on 33 independently generated data sets. The MCARMA$(1,0)$ model has no damping classification.}
\label{tab:param_recovery}
\begin{tabular}{llccc}
\hline\hline
Generative & Damping & \multicolumn{3}{c}{Nominal variability S/N}\\
\cline{3-5}
order & regime & $\approx10$ & $\approx4$ & $\approx2$\\
\hline
$(1,0)$ & ---         & 0.91 & 0.97 & 0.85\\
\hline
$(2,0)$ & Underdamped & 1.00 & 0.79 & 0.97\\
        & Critical    & 0.91 & 0.91 & 0.70\\
        & Overdamped  & 0.88 & 0.61 & 0.27\\
\hline
$(2,1)$ & Underdamped & 1.00 & 1.00 & 0.76\\
        & Critical    & 0.70 & 0.36 & 0.30\\
        & Overdamped  & 1.00 & 1.00 & 1.00\\
\hline
\end{tabular}
\end{table}

The first-order model is recovered reliably across all three noise levels, with correct-selection rates between 0.85 and 0.97. Recovery of the second-order models depends more strongly on their dynamical regime. For MCARMA$(2,0)$, the underdamped model is recovered well throughout the experiment, whereas recovery of the overdamped model decreases from 0.88 to 0.61 and 0.27 as variability S/N decreases. An overdamped second-order AR polynomial contains two real decay scales, and one component can become difficult to resolve when its characteristic timescale approaches the temporal range informed by the observing design.

This is an observational limitation rather than a property of the model: whether a characteristic timescale is resolved depends on the relationship between the dynamics and the observing window. To isolate it, we extend the MCARMA$(1,0)$ experiment over nominal characteristic timescales of 10, 30, 100, and 300 days while holding the model order and observing design fixed, so that no pole--zero cancellation is possible. Table~
\ref{tab:recovery.tauladder} shows that model-order recovery deteriorates gradually, whereas spectral recovery and the empirical stationarity diagnostic deteriorate much more rapidly as the characteristic timescale approaches the long-timescale boundary of the observations.

\begin{table}[!b]
\centering
\caption{Recovery and accuracy summaries for the extended MCARMA$(1,0)$ timescale experiment. Nominal timescales identify the centers of the generative ranges. ``Correct order'' is the proportion of data sets for which AICc selects MCARMA$(1,0)$; relative error and spectral error are medians over the corresponding setting; and ADF is the fraction classified as stationary in every band. Each setting contains 99 independent generative processes. The 300-day setting is a boundary case approaching the long-timescale limit of the observing design.}
\label{tab:recovery.tauladder}
\begin{tabular}{cccccc}
\hline\hline
Nominal $\tau$ & Drawn range & Correct & Rel.\ error & SNSE & ADF \\
(days) & (days) & order & $(\alpha_{1,1})$ & (all bands) & all-band\\
\hline
10  & 10.0--16.6   & 0.919 & 0.048 & 0.0146 & 1.00\\
30  & 23.3--38.6   & 0.909 & 0.060 & 0.0324 & 0.78\\
100 & 77.8--128.8  & 0.818 & 0.077 & 0.1311 & 0.18\\
\hline
\multicolumn{6}{l}{\emph{Boundary case}}\\
300 & 233--386      & 0.667 & 0.111 & 0.4744 & 0.02\\
\hline
\end{tabular}
\end{table}

Correct-order recovery decreases from 0.919 to 0.818 over the primary 10--100 day range and to 0.667 at the 300-day boundary case, but median SNSE increases from 0.0146 to 0.1311 and then 0.4744. The all-band stationarity-pass fraction falls even more sharply, from 1.00 to 0.18 and 0.02. Thus, a data set can contain sufficient information to select the appropriate stochastic order while providing much less information about the precise characteristic timescale or low-frequency PSD. Cross-band dependence can share information that is present in other bands, but it cannot create information about temporal scales unresolved in all of them. Full error distributions and additional diagnostics for this experiment are given in Appendix~\ref{sec:supp.simulation.identifiability}.

MCARMA$(2,1)$ exhibits a second form of weak identification that is structural rather than observational. The overdamped configuration is correctly selected in every simulation, whereas recovery of the critically damped configuration is only 0.70, 0.36, and 0.30 across the three S/N levels. Let
\[
 d_j=\min_{r,m}|\lambda_{r,j}-z_{m,j}|.
\]
denote the nearest pole--zero separation for band $j$. Pooling the 297 MCARMA\((2,1)\) data sets and dividing them into four equal-sized groups according to increasing pole--zero separation, correct-order recovery increases from 0.44 in the lowest-separation group to 0.81, 0.95, and 0.95 in successive groups, while the median natural frequency changes little across the groups. The deterioration is therefore associated primarily with pole--zero proximity. Near cancellation makes the transfer function insensitive to the separate locations of the pole and zero, so a nominal MCARMA$(2,1)$ process can become statistically close to a lower-order model even under otherwise informative observations.

\subsection{Benefits of joint multivariate estimation}
\label{sec:simulation.multivariate}

We next evaluate estimation accuracy at the generative model order and compare joint fitting of the five correlated bands with separate univariate CARMA fits. Because each band has the same marginal CARMA$(p,q)$ form under the two analyses, differences in accuracy isolate the information gained by modeling cross-band dependence rather than differences in the marginal stochastic model.

\begin{figure}[t]
\centering
\includegraphics[scale = 0.52]{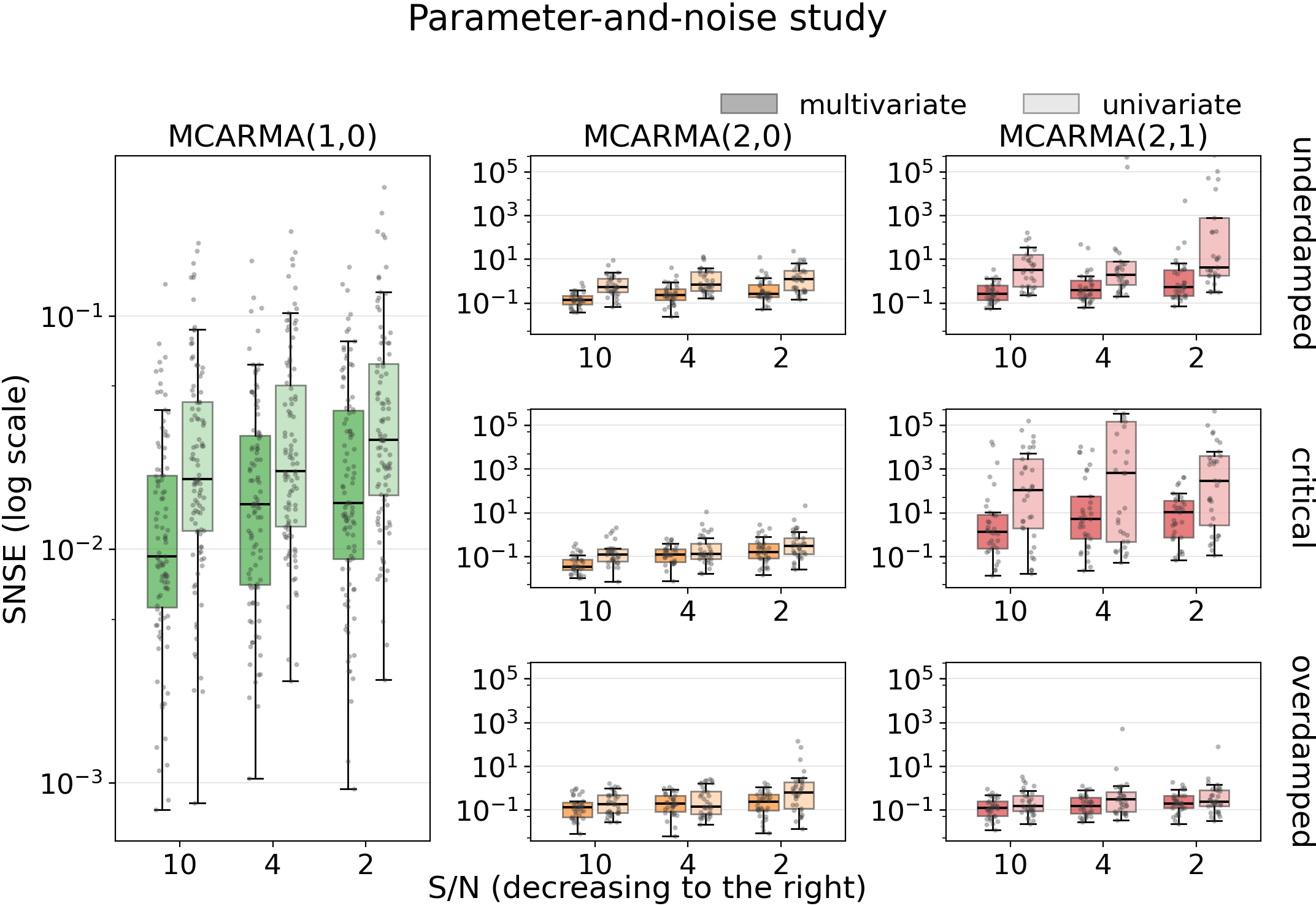}
\caption{Spectral accuracy of the joint multivariate MCARMA and separate univariate CARMA fits in the simulation study. Shown is the shape-normalized spectral error (SNSE) of the estimated marginal PSD at the generative order across the variability-S/N levels and damping regimes. Smaller values indicate more accurate spectral-shape recovery. The multivariate fit has smaller median SNSE in 26 of the 27 simulation settings.}
\label{fig:controlled_psd_accuracy}
\end{figure}

Figure~\ref{fig:controlled_psd_accuracy} compares SNSE under the multivariate and univariate fits. The multivariate fit has smaller median SNSE in 26 of the 27 simulation settings, and the median ratio of univariate to multivariate SNSE across settings is approximately 1.9. The advantage occurs at all three variability-S/N levels rather than being confined to the noisiest data sets. In several higher-order settings, separate univariate fits also produce spectral features that are reduced or absent under joint estimation, consistent with several correlated bands jointly constraining features that are only weakly supported by observations in an individual band.

Coefficient-wise recovery shows the same tendency. We use
$\alpha_{1,1}$ as a representative parameter because it is present
in all three model orders considered here. Joint fitting has smaller
relative error in $\alpha_{1,1}$ in 23 of 27 settings, with a median
univariate-to-multivariate error ratio of approximately 1.3.
Appendix~\ref{sec:supp.simulation.multivariate} reports the corresponding coefficient-wise
results and additional spectral-accuracy analyses, which support the
same qualitative conclusion about the benefit of joint fitting.

The larger gain for spectral recovery than for a single coefficient has a natural interpretation. The marginal PSD is determined jointly by the AR and MA parameters and stochastic driving amplitude, so modest improvements in several estimated quantities can combine into a substantially more accurate estimate of the stochastic spectrum. Conversely, individual coefficients can be weakly determined while different parameter combinations imply similar spectra over the frequencies directly informed by the observations. The benefit of multivariate modeling is therefore not restricted to estimation of cross-band dependence: correlated bands also provide information about one another's marginal stochastic dynamics.

\begin{figure}[b]
\centering
\includegraphics[scale = 0.9]{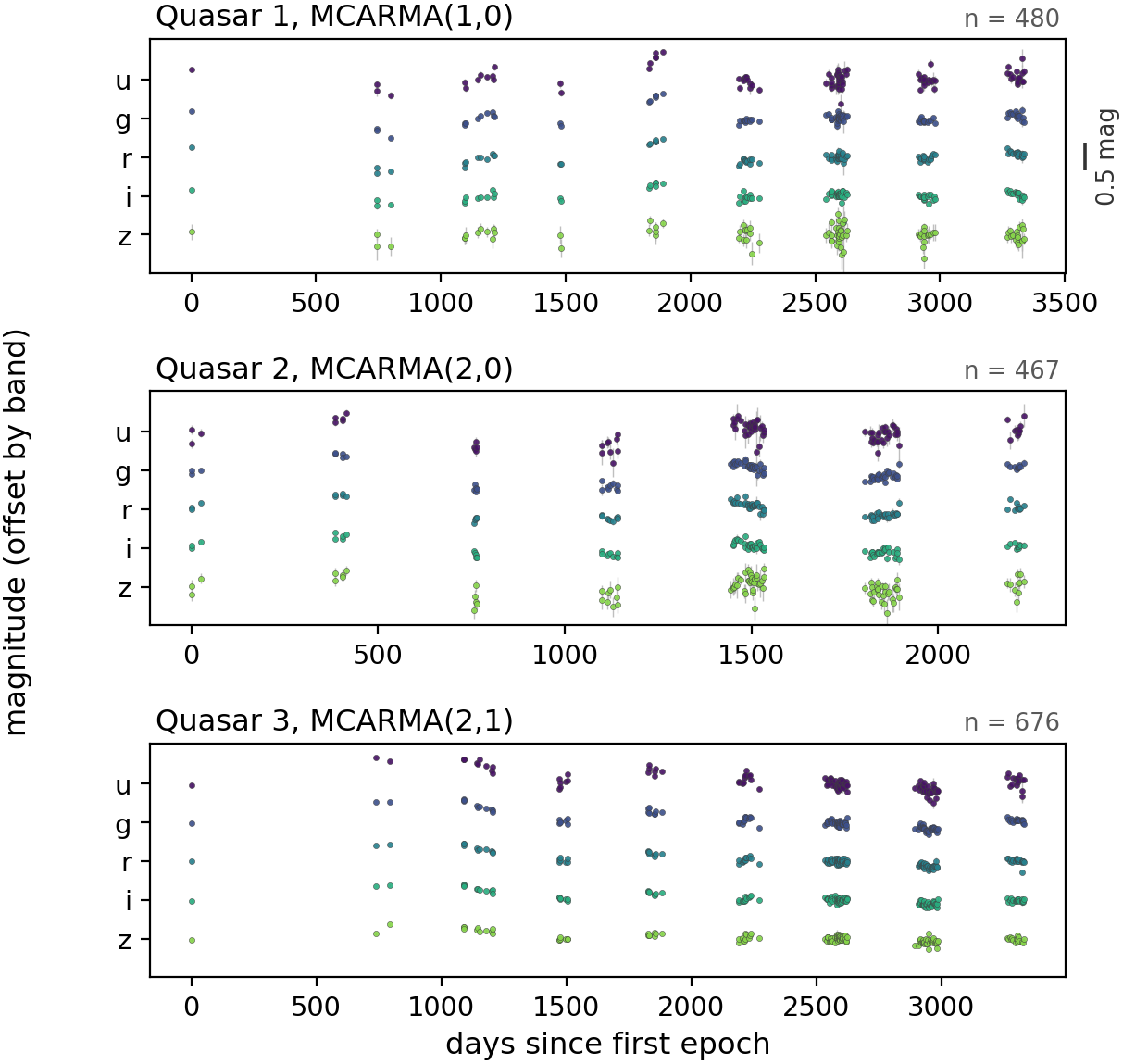}
\caption{Observed five-band $ugriz$ data for Quasars~1--3, one
panel per object. Points are the measured magnitudes and vertical bars
indicate the reported $1\sigma$ measurement uncertainties. Following
the astronomical convention, magnitude increases toward fainter
sources. Within each panel the five bands are centered on their own
medians and offset vertically by a constant amount, so the vertical
axis is labeled by band; the scale bar gives the magnitude scale,
which is common to all three panels. Time is measured from the first
epoch, and the total number of observations is annotated in each
panel. The seasonal gaps are evident in all three objects; Quasar~2
has the shortest baseline, whereas Quasar~3 has the largest number of
observations and the highest stationary signal-to-noise ratios.}
\label{fig:sdss.data}
\end{figure}

Appendices~\ref{sec:supp.simulation.uncertainty}--\ref{sec:supp.simulation.computation} further show that likelihood-curvature uncertainty is reasonably calibrated in well-identified settings but becomes less reliable for higher-order and weakly identified configurations; Hessian standard errors should therefore be interpreted as local uncertainty summaries rather than uniformly calibrated finite-sample confidence intervals. Sensitivity analyses show that the regularized first stage lies very close to the ordinary-likelihood maximum for correctly specified models, while the second-stage refinement ensures numerical regularization does not enter the final model comparison. A sensitivity analysis using the generalized information criterion of \citet{10.1093/biomet/76.2.297} yields lower overall correct-order recovery than AICc (0.671 versus 0.827), supporting AICc as the primary criterion in these simulations.
Additional RTS reconstruction experiments are reported in
Appendix~\ref{sec:supp.simulation.rts}, while computational benchmarks are given
in Appendix~\ref{sec:supp.simulation.computation}; complete five-band fits require
approximately 147 seconds for MCARMA$(1,0)$ and 792--867 seconds for
the second-order models under the production optimization procedure.

\section{SDSS Stripe 82 Quasar Case Studies}
\label{sec:sdss}

We illustrate the proposed methodology using five-band time series of
three quasars observed by the Sloan Digital Sky Survey (SDSS).
Whereas the simulations of Section~\ref{sec:simulation} permit
statistical performance to be evaluated against known generative
processes, these case studies illustrate how fitted MCARMA dynamics
and their spectral representations can be interpreted when the
underlying stochastic process is unknown. We deliberately choose one
object for which AICc selects each of MCARMA$(1,0)$,
MCARMA$(2,0)$, and MCARMA$(2,1)$. The examples are therefore
illustrative and are not intended to characterize the relative
frequency of the three orders in the quasar population.

\subsection{Data and analysis}
\label{sec:sdss.data}

We consider the SDSS Stripe~82 five-band \(ugriz\) time series of
9,258 spectroscopically confirmed quasars from the SDSS Data Release 7
quasar catalog compiled by \citet{macleod2012}. Requiring at least 400
individual-band measurements across the five filters leaves 386
candidate objects. After linear detrending, a per-band stationarity
screen leaves 134 objects compatible with the stationary stochastic
models considered here. For each object, we fit MCARMA$(1,0)$,
MCARMA$(2,0)$, and MCARMA$(2,1)$ using the two-stage procedure of
Section~\ref{sec:estimation} and select the preferred model order by
AICc. From the 134 eligible objects, we select three examples with
particularly strong AICc preference for MCARMA$(1,0)$,
MCARMA$(2,0)$, and MCARMA$(2,1)$, respectively, denoted
by Quasars~1--3.

For comparison with the band-by-band analysis commonly used in
astronomical applications, we also fit each band independently,
selecting its univariate CARMA order separately by AICc from
CARMA$(1,0)$, CARMA$(2,0)$, and CARMA$(2,1)$. These independently
selected fits therefore represent a complete single-band analysis
rather than fits conditioned on the order selected by the joint
MCARMA analysis. Figure~\ref{fig:sdss.data} shows the five-band
observations, and Table~\ref{tab:sdss.summary} summarizes their
observing characteristics and principal fitted features. Complete
parameter estimates, curvature-based standard errors, and diagnostics
identifying fitted timescales outside the range constrained by each
observing design are reported in Appendix~\ref{sec:supp.sdss}.

\begin{table}[t]
\centering
\caption{Summary of the three SDSS quasar case studies. $N$ is the
total number of individual-band measurements,
$\Delta\mathrm{AICc}$ is the difference between the selected and
next-best models, and S/N is the range of stationary signal-to-noise
ratios across the five bands.}
\label{tab:sdss.summary}
\begin{tabular}{lccc}
\hline\hline
 & Quasar~1 & Quasar~2 & Quasar~3 \\
\hline
Catalog ID            & 1995955 & 1465040 & 3064008 \\
$N$                   & 480 & 467 & 676 \\
Baseline (days)       & 3337 & 2230 & 3331 \\
Median spacing (days) & 3.0 & 2.0 & 2.0 \\
Stationary S/N        & 0.6--7.0 & 1.2--8.0 & 2.7--7.9 \\
Selected order        & $(1,0)$ & $(2,0)$ & $(2,1)$ \\
$\Delta\mathrm{AICc}$ & 5.55 & 8.30 & 215.13 \\
Timescales (days)     & 374--493; slow & 7--25; fast & 430--540; slow \\
High-frequency PSD    & $f^{-2}$ & $f^{-4}$ & $f^{-2}$ with MA flattening \\
Coherence             & $z$-band pairs less precise
                      & non-monotone
                      & decreases with separation \\
\hline
\end{tabular}
\end{table}

\subsection{Three contrasting stochastic structures}
\label{sec:sdss.cases}

Quasar~1 has 480 measurements over 3337 days, with six seasonal
observing groups and a median within-band spacing of 3.0 days. Its
$z$ band is comparatively weak: the fitted stationary standard
deviation divided by the median reported measurement uncertainty is
only 0.6, compared with 2.6--7.0 in the other bands. AICc selects
MCARMA$(1,0)$, but only by 5.55 over MCARMA$(2,0)$; the appropriate
interpretation is therefore that these data do not require
higher-order structure rather than that they strongly exclude it.
Under the selected model the five decay timescales range from 374 to
493 days and all lie within the interval constrained by the observing
design.

\begin{figure}[p]
\centering
\includegraphics[scale=0.88]{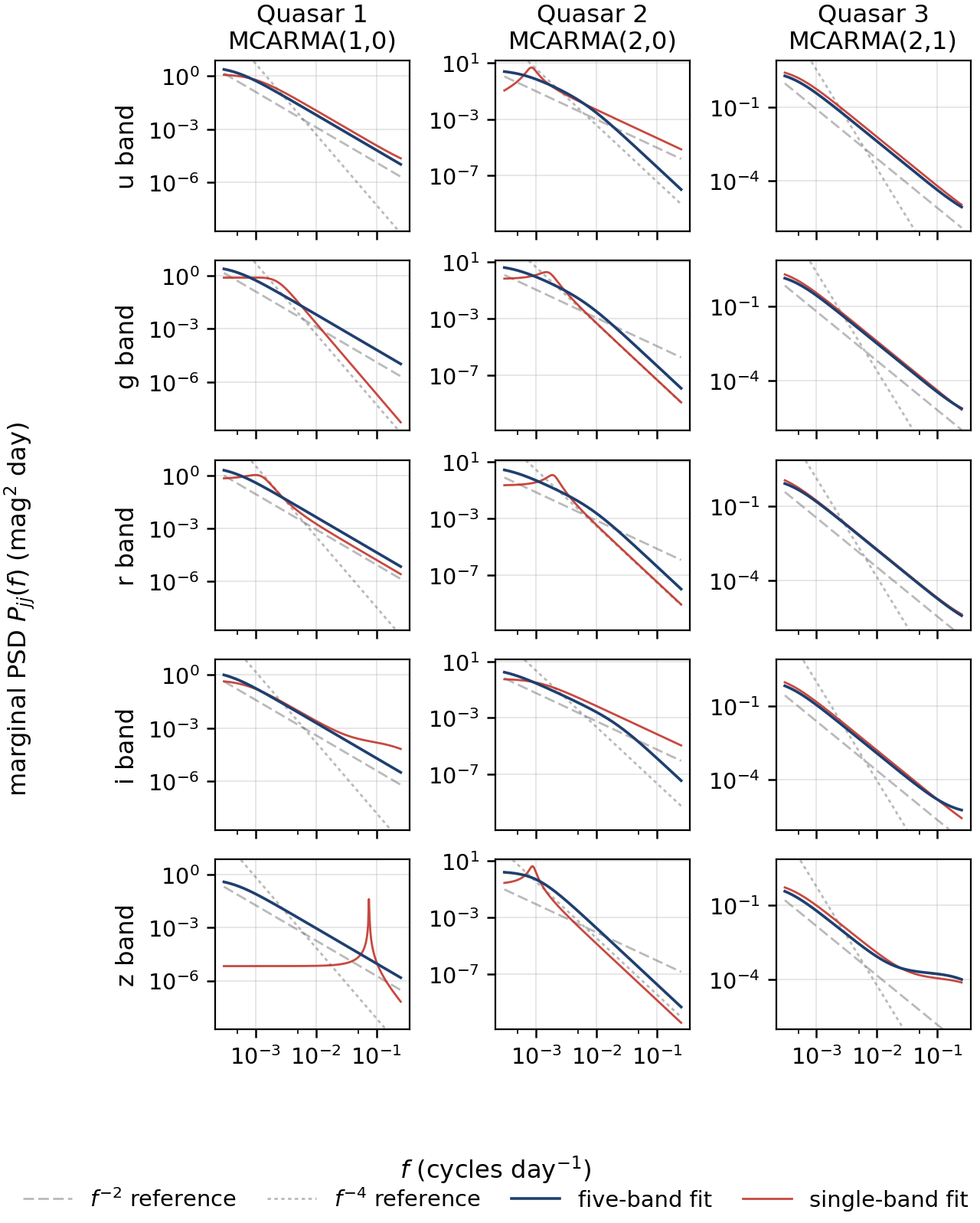}
\caption{Estimated marginal PSDs for Quasars~1--3. Blue curves are
the marginal PSDs from the AICc-selected joint five-band MCARMA model.
Red curves are obtained by fitting each band independently, with the
univariate CARMA order selected separately by AICc from
CARMA$(1,0)$, CARMA$(2,0)$, and CARMA$(2,1)$. Thus, the red curves
need not have the same model order as the corresponding joint MCARMA
fit. Dashed and dotted reference lines indicate $f^{-2}$ and $f^{-4}$
behavior. The comparison illustrates how joint multiband analysis can
affect both model-order selection and the resulting marginal spectral
inference.}
\label{fig:sdss.psd}
\end{figure}

The corresponding joint marginal PSDs in
Figure~\ref{fig:sdss.psd} are approximately flat below their
characteristic frequencies and then decline toward the $f^{-2}$
high-frequency behavior of a first-order AR process. The independently
selected single-band fits give a different picture: several exhibit
higher-order spectral structure, most prominently the pronounced
finite-frequency feature in the $z$ band. Thus, the joint analysis
supports a common first-order stochastic order with band-specific
dynamics, whereas separate analysis can lead to substantially
different stochastic descriptions across bands of the same quasar.
The contrast is particularly notable for the weak $z$ band and
illustrates how joint multiband modeling can use cross-band dependence
to inform marginal stochastic dynamics when an individual band is
only weakly informative. Because the underlying stochastic process is
unknown, however, this comparison does not establish which
representation is closer to the truth.

Quasar~2 contains 467 measurements over a shorter 2230-day baseline.
AICc selects MCARMA$(2,0)$ by 8.30 over MCARMA$(1,0)$ and 10.67 over
MCARMA$(2,1)$, supporting a second AR pole without requiring an MA
zero. The fitted fast timescales are approximately 7--25 days in four
bands and are much longer than the two-day median sampling interval.
Some slow-timescale estimates, however, extend beyond the practical
upper range implied by the baseline, illustrating that model-order
selection does not imply that every fitted timescale is equally well
constrained.

The joint marginal spectra show the much steeper $f^{-4}$
high-frequency decline implied by two autoregressive poles with no MA
zero. In contrast, the independently selected single-band fits produce
substantially different spectral shapes across bands, including
finite-frequency features absent from the joint fit. The joint
analysis therefore supports a common second-order AR structure with
band-specific dynamics, whereas separate analysis leads to more
heterogeneous stochastic descriptions across bands. This comparison
illustrates how sharing information across dependent bands can
materially affect both model-order selection and marginal spectral
inference. As in Quasar~1, the true stochastic structure is unknown,
so the comparison does not establish which description is closer to
the truth.

Quasar~3 is the best observed of the three, with 676 measurements over
3331 days and stationary S/N ratios between 2.7 and 7.9. AICc selects
MCARMA$(2,1)$ decisively: its AICc is lower than that of the next-best
model by 215.13. Moreover, MCARMA$(2,0)$ performs worse than
MCARMA$(1,0)$, indicating that the improvement is associated
specifically with the MA component rather than merely with adding AR
order. Consistent with this interpretation, the five fitted MA
coefficients are each separated from zero by approximately three to
four local standard errors.

The fitted slow timescales are approximately 430--540 days. In
contrast, the fitted fast AR timescales, 0.003--0.585 days, lie far
below the two-day sampling scale, and several MA-zero timescales are
also near or below the cadence bound. We therefore do not interpret
the fast pole and MA zero as separately resolved physical timescales.
Their scientifically interpretable consequence is instead the
modification they induce in the short-timescale spectrum. Over the
plotted frequency range the fitted PSDs show a shallower decline than
the purely AR cases, including visible flattening in the $i$ and $z$
bands. This is a finite-frequency effect of the MA zero; the
asymptotic MCARMA$(2,1)$ slope remains $f^{-2}$. In contrast to the
first two quasars, the joint and independently selected single-band
fits agree closely in several bands. This example therefore shows
that joint fitting need not materially alter marginal spectral
inference when the individual-band analyses already support similar
spectral structure.

\subsection{Cross-band dependence}
\label{sec:sdss.coherence}

The marginal PSDs describe band-specific stochastic dynamics, whereas
the multivariate model additionally estimates cross-band dependence.
Under the diagonal dynamic structure of Section~\ref{sec2}, the
coherence between two bands equals the squared correlation of their
stochastic drivers and is constant with frequency.
Table~\ref{tab:coherence} therefore provides a complete summary of
this aspect of the fitted dependence.

\begin{table}[b]
\centering
\caption{Estimated pairwise magnitude-squared coherence for the three
case studies, each evaluated under its AICc-selected model. Standard
errors in parentheses are obtained by the delta method through the
Cholesky factor of the driving covariance. Coherence is
frequency-independent under the adopted diagonal dynamic structure.
An estimate of 1.000 lies on the boundary of the correlation space.
Its near-zero standard error should therefore not be interpreted as
high precision.}
\label{tab:coherence}
\begin{tabular}{lrrrr}
\hline\hline
 & $u$ & $g$ & $r$ & $i$ \\
\hline
\multicolumn{5}{l}{Quasar~1, MCARMA$(1,0)$} \\
$g$ & 1.000 (0.000) &  &  &  \\
$r$ & 0.974 (0.022) & 0.974 (0.022) &  &  \\
$i$ & 0.993 (0.010) & 0.993 (0.010) & 0.994 (0.008) &  \\
$z$ & 0.979 (0.053) & 0.979 (0.053) & 0.909 (0.120) & 0.949 (0.089) \\
\hline
\multicolumn{5}{l}{Quasar~2, MCARMA$(2,0)$} \\
$g$ & 0.999 (0.003) &  &  &  \\
$r$ & 0.967 (0.040) & 0.975 (0.022) &  &  \\
$i$ & 0.961 (0.047) & 0.970 (0.029) & 1.000 (0.001) &  \\
$z$ & 1.000 (0.004) & 1.000 (0.002) & 0.972 (0.045) & 0.967 (0.052) \\
\hline
\multicolumn{5}{l}{Quasar~3, MCARMA$(2,1)$} \\
$g$ & 1.000 (0.000) &  &  &  \\
$r$ & 0.996 (0.006) & 0.996 (0.006) &  &  \\
$i$ & 0.973 (0.027) & 0.973 (0.027) & 0.990 (0.012) &  \\
$z$ & 0.951 (0.060) & 0.951 (0.060) & 0.976 (0.037) & 0.997 (0.011) \\
\hline
\end{tabular}
\end{table}

All three objects exhibit strong fitted cross-band dependence, with
coherence estimates between 0.909 and 1.000. Quasar~1 has larger
uncertainty for pairs involving the comparatively weak $z$ band.
Quasar~3 shows the clearest decrease in dependence with increasing
wavelength separation, whereas Quasar~2 has no monotone wavelength
ordering. The fitted driving covariance is nearly rank one in all
three cases, with its leading component accounting for
99.0--99.2\% of the driving variation. Near-unit coherence estimates
should therefore be interpreted cautiously because a near-singular
driving covariance can push estimated correlations toward the
boundary; the corresponding parameter and curvature diagnostics are
given in Appendix~\ref{sec:supp.sdss.identifiability}.

Taken together, the three examples illustrate complementary aspects
of MCARMA inference from real multi-band observations. The joint
model selects a common stochastic order while retaining band-specific
dynamics, whereas independently fitted bands can favor different
orders and spectral structures. This distinction is substantial for
Quasars~1 and 2 but less pronounced for Quasar~3. The examples also
show why fitted parameters must be interpreted relative to the
temporal resolution of the observations: all fitted timescales of
Quasar~1 are within the informative range, some slow timescales of
Quasar~2 extend beyond its baseline constraint, and the fast pole of
Quasar~3 lies well below its cadence. Finally, the strong fitted
coherence shows that the bands contain substantial cross-band
dependence unavailable to separate single-band analyses. Joint
multivariate modeling can use this dependence to inform marginal
stochastic dynamics when individual bands are only partially
informative.

\section{Conclusion}
\label{sec:conclusion}

We have developed a structured subclass of Wiener-driven Gaussian
multivariate continuous-time autoregressive moving average (MCARMA)
processes for irregularly sampled multi-band time series. By combining
component-specific CARMA dynamics with correlated Brownian drivers,
the model separates marginal temporal behavior from cross-band
dependence while retaining a state-space representation for
likelihood-based computation. Simulation studies show that joint
multivariate estimation can substantially improve recovery over
separate single-band analyses, particularly for spectral quantities
that summarize the combined effects of several dynamical parameters.
The SDSS Stripe~82 examples further illustrate an important consequence
of multiband modeling: joint MCARMA estimation can use cross-band
dependence to inform marginal stochastic dynamics while retaining
band-specific parameters. For the first two quasars, the joint and
independently selected single-band models yield substantially different
model-order and spectral inference, whereas the two approaches agree
more closely for the third. Because the underlying stochastic
processes are unknown, these comparisons demonstrate the impact of
joint versus separate analysis rather than establishing which
representation is closer to the truth.

An important lesson from the simulations is that model selection,
parameter identifiability, and recovery of observable stochastic
behavior are related but distinct questions. AICc compares candidate
models according to their fit to the observed data while penalizing
model complexity; it does not require every parameter of the selected
model to be estimated with equal precision. Consequently, selection
of a higher-order model need not imply that every fitted pole, zero,
or associated timescale is separately resolved by the observations.
A characteristic timescale may be weakly constrained when it approaches
or lies beyond the temporal range effectively probed by the cadence
and baseline, even when the fitted higher-order process has
distinguishable spectral behavior over the informative frequency
range. The simulations show that the marginal PSD can remain well
recovered in such cases.

Near pole--zero cancellation presents a different limitation. As an
MA zero approaches an AR pole, their effects increasingly cancel and
the higher-order process itself becomes difficult to distinguish from
a lower-order process. This is therefore not merely uncertainty in an
individual timescale but loss of information about the additional
stochastic structure. Model-order selection and parameter
interpretation should accordingly be considered together with
diagnostics of observational resolution and pole--zero separation.
In particular, selection of a higher-order model should not by itself
be interpreted as evidence that each fitted pole or zero represents a
separately resolved physical timescale. Functionals such as the power
spectral density can provide more stable summaries of the stochastic
behavior actually constrained by the observations.

The structured model considered here also suggests several directions
for extending multivariate continuous-time modeling. We deliberately
use diagonal AR and MA coefficient matrices and introduce dependence
through correlated Brownian drivers. This yields a parsimonious
separation between marginal dynamics and cross-component dependence,
but also implies frequency-independent coherence. Allowing selected
off-diagonal dynamic terms, alternative latent driving structures, or
other structured matrix parameterizations could accommodate
frequency-dependent cross-spectral behavior while avoiding the large
number of parameters in an unrestricted MCARMA model. Determining
which such structures remain identifiable under irregular and
partially observed sampling is an important statistical problem in
its own right.

There is also considerable scope for developing inference beyond the
likelihood optimization considered here. Weakly identified
higher-order models can place covariance estimates near the boundary
of the positive-definite cone, where local curvature-based uncertainty
quantification becomes unreliable. Bayesian inference could propagate
these uncertainties more fully, although computation and prior
specification become increasingly challenging as the model order and
number of components grow. Simulation-based inference and other
amortized approaches
\citep{2020PNAS..11730055C,papamakarios2019}
may provide useful alternatives when the same model must be fitted to
large ensembles of irregular multivariate time series. Hierarchical
extensions are another natural direction, particularly when many
related objects are observed and information about their stochastic
dynamics can be shared at the population level.

Although motivated by astronomical multi-band observations, the
statistical issues addressed here are not specific to astronomy.
Irregular and partially observed multivariate measurements arise in
many settings where continuous-time dependence is scientifically
meaningful. The structured MCARMA construction provides one way to
balance flexible component-specific dynamics, interpretable
cross-component dependence, and computational tractability. More
broadly, the results show that the value of multivariate modeling is
not limited to estimating dependence itself: cross-component
dependence provides information that can be used to infer marginal
stochastic dynamics when individual series are only partially
informative. The publicly available \texttt{mcarma} package
\citep{mcarma2026} provides a basis for exploring these extensions
and applications.

\begin{acks}[Acknowledgments]
The corresponding author, H. Tak,  gratefully acknowledges partial support from the Brainpool Fellowship Program 2026–27 of the National Research Foundation in South Korea (RS-2026-25551883).
\end{acks}

%%%%%%%%%%%%%%%%%%%%%%%%%%%%%%%%%%%%%%%%%%%%%%
%% Funding information, if any,             %%
%% should be provided in the                %%
%% funding section.                         %%
%%%%%%%%%%%%%%%%%%%%%%%%%%%%%%%%%%%%%%%%%%%%%%
%\begin{funding}
%The first author was supported by NSF Grant DMS-??-??????.

%The corresponding author, H. Tak,  was supported in part by the Brainpool Fellowship Program 2026–27 of the National Research Foundation in South Korea (RS-2026-25551883).
%\end{funding}

\appendix

% The journal supplementary material is incorporated below as appendices
% for the arXiv version. Cross-references have been converted to live links.

\section{Supplementary Methodology}
\label{sec:supp.method}

These appendices provide derivations and implementation details for
Sections~\ref{sec2}--\ref{sec:estimation}. The main
text contains the model definition, likelihood, spectral quantities,
root-based interpretation, two-stage estimator, uncertainty
quantification, and AICc model-selection rule required to interpret
the numerical studies. Here we collect the equivalent covariance
representation, complete Kalman recursions, optimization
parameterization, numerical regularization, and convergence
diagnostics.

\subsection{State-space and covariance representations}
\label{sec:supp.statespace}

The representation used here is the controller canonical form, which
stacks the derivatives of a latent process rather than of
$\mathbf{X}(t)$ itself. Let $\mathbf{U}(t)$ be the $k$-dimensional
process driven by the autoregressive operator alone, so that
$A(D)\mathbf{U}(t)=d\mathbf{B}(t)$ with $A(\cdot)$ as in the main text,
and define
\[
\mathbf{Z}(t)
=
\left[
\mathbf{U}(t)^\top,\,
D\mathbf{U}(t)^\top,\,
\ldots,\,
D^{p-1}\mathbf{U}(t)^\top
\right]^\top.
\]
The observed process is recovered by applying the moving-average
operator, $\mathbf{X}(t)=M(D)\mathbf{U}(t)=\sum_{l=0}^{q}M_lD^l\mathbf{U}(t)$
with $M_0=I_k$, which is exactly the linear map $H$ given below.
Because $A(\cdot)$ and $M(\cdot)$ have diagonal coefficient matrices here
they commute, so $\mathbf{X}$ satisfies $A(D)\mathbf{X}(t)=M(D)\,
d\mathbf{B}(t)$ as required. The state vector reduces to the stack of
derivatives of $\mathbf{X}(t)$ itself only when $q=0$, in which case
$M(D)=I_k$ and $\mathbf{U}(t)=\mathbf{X}(t)$.

The MCARMA state-space representation is
\[
d\mathbf{Z}(t)=F\mathbf{Z}(t)\,dt+G\,d\mathbf{B}(t),
\qquad
\mathbf{X}(t)=H\mathbf{Z}(t),
\]
with
\[
F=
\begin{bmatrix}
0_k&I_k&0_k&\cdots&0_k\\
0_k&0_k&I_k&\cdots&0_k\\
\vdots&&&\ddots&\vdots\\
-A_0&-A_1&-A_2&\cdots&-A_{p-1}
\end{bmatrix},
\qquad
G=
\begin{bmatrix}
0_k\\
\vdots\\
0_k\\
I_k
\end{bmatrix},
\]
and
\[
H=
\begin{bmatrix}
M_0&M_1&\cdots&M_{p-1}
\end{bmatrix},
\qquad
M_0=I_k,\quad M_l=0_k\quad\text{for }l>q.
\]
The superdiagonal blocks of $F$ are $+I_k$ throughout, so successive
block components of $\mathbf{Z}(t)$ are ordinary derivatives of
$\mathbf{U}(t)$ with no sign alternation. The minus signs appear only
in the final block row, which encodes
$D^{p}\mathbf{U}(t)=-\sum_{j=0}^{p-1}A_jD^{j}\mathbf{U}(t)+d\mathbf{B}(t)/dt$.
Under the diagonal dynamic structure, a permutation of the state
coordinates separates $F$ into band-specific companion systems, while
their innovations remain coupled through $V$.

For a stationary solution, the state covariance $P_\infty$ solves
\[
FP_\infty+P_\infty F^\top+GVG^\top=0,
\]
equivalently
\[
P_\infty
=
\int_0^\infty
e^{Fs}GVG^\top e^{F^\top s}\,ds.
\]
The process covariance is
\[
C(h)
=
\begin{cases}
He^{Fh}P_\infty H^\top,&h\geq0,\\
HP_\infty e^{-F^\top h}H^\top,&h<0,
\end{cases}
\]
with $C(-h)=C(h)^\top$. Hence
\[
\mathbf{X}(t)\sim\mathcal{GP}\{0,C(t-t')\}.
\]
At times $t_1,\ldots,t_n$, the stacked latent process is Gaussian with
a block covariance matrix whose $(i,j)$ block is $C(t_i-t_j)$.
Consequently, the covariance-based and state-space formulations define
the same finite-dimensional Gaussian distributions and the same
likelihood. The state-space form avoids factorization of the full
dense covariance matrix.

For the MCARMA$(1,0)$ special case, write the band-specific decay rate
as $\lambda_j=1/\tau_j$ and
$V_{jl}=\sigma_j\sigma_l\rho_{jl}$. Then
\[
\operatorname{Cov}\{X_j(t),X_l(t)\}
=
\frac{\sigma_j\sigma_l\rho_{jl}}{\lambda_j+\lambda_l}
=
\sigma_j\sigma_l\rho_{jl}
\frac{\tau_j\tau_l}{\tau_j+\tau_l},
\]
and
\[
\operatorname{Var}\{X_j(t)\}
=
\frac{\sigma_j^2\tau_j}{2}.
\]

\subsection{Additional spectral properties}
\label{sec:supp.spectral}

The spectral density can equivalently be written in state-space form as
\[
P(\omega)
=
H(i\omega I_{kp}-F)^{-1}
GVG^\top
(-i\omega I_{kp}-F^\top)^{-1}
H^\top.
\]
For a real-valued process,
\[
P(-\omega)=\overline{P(\omega)}=P(\omega)^\top.
\]
The real and imaginary parts of an off-diagonal element
$P_{jl}(\omega)$ are the cospectrum and quadrature spectrum,
respectively.

Although coherence is constant under the diagonal dynamic structure,
cross-spectral phase can vary with frequency. If
\[
h_j(i\omega)=|h_j(i\omega)|e^{i\phi_j(\omega)}
\]
and $V_{jl}$ is real, then, up to the sign of $V_{jl}$,
\[
\arg\{P_{jl}(\omega)\}
=
\phi_j(\omega)-\phi_l(\omega).
\]
This phase difference should not in general be interpreted as a
physical inter-band delay: the model contains no explicit delay
parameter, and the phase arises from different component-specific
linear filters acting on correlated stochastic drivers.

The high-frequency marginal behavior follows from the degrees of the
AR and MA polynomials:
\[
P_{jj}(\omega)
=
O\!\left(|\omega|^{-2(p-q)}\right).
\]
In particular, MCARMA$(1,0)$ and MCARMA$(2,1)$ have asymptotic slope
$-2$, while MCARMA$(2,0)$ has asymptotic slope $-4$. Finite-frequency
behavior can differ substantially from these asymptotic slopes when
poles or zeros lie near the observed frequency range.

\subsection{Exact irregular-time transitions and Kalman recursions}
\label{sec:supp.kalman}

For $\Delta_i=t_i-t_{i-1}$,
\[
\mathbf{Z}(t_i)
=
e^{F\Delta_i}\mathbf{Z}(t_{i-1})
+
\int_{t_{i-1}}^{t_i}
e^{F(t_i-s)}G\,d\mathbf{B}(s).
\]
Thus
\[
\Phi_i=e^{F\Delta_i},
\qquad
Q_i
=
\int_0^{\Delta_i}
e^{Fs}GVG^\top e^{F^\top s}\,ds.
\]
Under stationarity,
\[
Q_i=P_\infty-\Phi_iP_\infty\Phi_i^\top,
\]
which is the identity used in computation. This avoids direct
evaluation of the integral and is convenient because $P_\infty$ is
already needed for stationary initialization.

Let $\mathbf{m}_{i|j}$ and $P_{i|j}$ denote the conditional state mean
and covariance at $t_i$ given observations through $t_j$. Initialize
\[
\mathbf{m}_{1|0}=0,
\qquad
P_{1|0}=P_\infty.
\]
For $i\geq2$,
\begin{align*}
\mathbf{m}_{i|i-1}
&=\Phi_i\mathbf{m}_{i-1|i-1},\\
P_{i|i-1}
&=\Phi_iP_{i-1|i-1}\Phi_i^\top+Q_i.
\end{align*}
With innovation $\mathbf{x}_i^\ast$ and covariance $W_i^\ast$ as
defined in the main text, the Kalman gain is
\[
K_i
=
P_{i|i-1}H^\top S_i^\top(W_i^\ast)^{-1},
\]
and
\[
\mathbf{m}_{i|i}
=
\mathbf{m}_{i|i-1}
+
K_i\mathbf{x}_i^\ast.
\]
For numerical stability, the covariance is updated in Joseph form,
\[
P_{i|i}
=
(I_{kp}-K_iS_iH)P_{i|i-1}(I_{kp}-K_iS_iH)^\top
+
K_iR_i^\ast K_i^\top.
\]

The selection-matrix formulation handles arbitrary observation
patterns. For example, if $k=6$ and only components 2 and 4 are
observed at $t_i$,
\[
S_i=
\begin{bmatrix}
0&1&0&0&0&0\\
0&0&0&1&0&0
\end{bmatrix}.
\]
Observations at exactly the same time are processed jointly; distinct
times are processed sequentially even when closely spaced. No
interpolation or artificial synchronization is required.

With straightforward dense linear algebra, the dominant operations
are bounded by
\[
O\!\left(
n(kp)^3+\sum_{i=1}^{n}(k_i^\ast)^3
\right).
\]
For fixed $k,p,q$, this is linear in the number of unique observation
times. By contrast, direct evaluation of the equivalent dense
Gaussian-process likelihood for
$N=\sum_i k_i^\ast$ scalar observations requires $O(N^3)$ computation
and $O(N^2)$ storage.

The same state-space model supports Rauch--Tung--Striebel smoothing
\citep{rauch1965maximum}. The smoother combines the forward filtering
distributions with later observations to obtain conditional latent
state distributions between and at observed epochs.

\subsection{Optimization parameterization}
\label{sec:supp.parameterization}

For each component, the order-$p$ AR polynomial is parameterized as
products of stable linear and quadratic factors,
\begin{align*}
A_j(z)
={}&
(a_{1,j}+a_{2,j}z+z^2)
(a_{3,j}+a_{4,j}z+z^2)\cdots\\
&\times
\begin{cases}
a_{p-1,j}+a_{p,j}z+z^2,&p\ \text{even},\\
a_{p,j}+z,&p\ \text{odd},
\end{cases}
\end{align*}
with $a_{i,j}>0$. Similarly, under $M_j(0)=1$,
\begin{align*}
M_j(z)
={}&
(1+b_{1,j}z+b_{2,j}z^2)
(1+b_{3,j}z+b_{4,j}z^2)\cdots\\
&\times
\begin{cases}
1+b_{q-1,j}z+b_{q,j}z^2,&q\ \text{even},\\
1+b_{q,j}z,&q\ \text{odd},
\end{cases}
\end{align*}
with $b_{i,j}>0$. Positivity places all roots in the open left
half-plane. The models used in the paper have $p\leq2$ and $q\leq1$,
so no factor-ordering convention is required.

The same polynomials can be written as
\[
A_j(z)=\prod_{r=1}^{p}(z-\lambda_{r,j}),
\qquad
M_j(z)=c_j\prod_{m=1}^{q}(z-z_{m,j}),
\]
where $c_j$ is determined by $M_j(0)=1$. The factor coefficients are
used for optimization, while poles and zeros are derived for
interpretation and regularization.

We optimize over
\[
\widetilde a_{i,j}=\log a_{i,j},
\qquad
\widetilde b_{i,j}=\log b_{i,j},
\]
so the search is unconstrained while stationarity and minimum phase
are maintained. The minimum-phase restriction is an identifiability
convention: the marginal PSD identifies the magnitude of the transfer
function but not an unrestricted MA phase, so the convention selects
a representative among second-order-equivalent MA descriptions.

For the driving covariance,
\[
V=LL^\top,
\]
with
\[
L_{jj}=\exp(\theta_{V,jj}/2),
\qquad
L_{jl}=\theta_{V,jl}\quad(j>l).
\]
This guarantees positive definiteness. It does not prevent $V$ from
approaching singularity, which motivates the first-stage numerical
stabilization below.

For MCARMA$(2,0)$, the damping representation
\[
A_j(z)=z^2+2\zeta_j\omega_{n,j}z+\omega_{n,j}^2
\]
also shows that a nonzero-frequency PSD maximum occurs only when
\[
\zeta_j<\frac{1}{\sqrt{2}},
\qquad
\omega_{\rm peak,j}
=
\omega_{n,j}\sqrt{1-2\zeta_j^2}.
\]
Thus underdamped dynamics do not by themselves imply a spectral peak.

\subsection{Pole--zero cancellation}
\label{sec:supp.identifiability}

If an AR pole and MA zero coincide, write
\[
A_j(z)=(z-\lambda)A_j^\dagger(z),
\qquad
M_j(z)=(z-\lambda)M_j^\dagger(z).
\]
Then
\[
\frac{M_j(z)}{A_j(z)}
=
\frac{M_j^\dagger(z)}{A_j^\dagger(z)},
\]
so the nominal higher-order representation is nonminimal. For
MCARMA$(2,1)$ the MA zero is real under our parameterization; exact
cancellation can therefore occur only with a real AR pole. Critically
damped and overdamped dynamics can admit exact cancellation, whereas
a real MA zero cannot exactly cancel one member of a genuinely
complex-conjugate underdamped pair. Near cancellation produces a
nearly flat likelihood direction even when the implied transfer
function or PSD remains well determined.

\subsection{Regularized initialization}
\label{sec:supp.regularization}

The first-stage optimization uses soft regularization to avoid
numerically problematic regions while constructing an initialization
for the ordinary likelihood. For a positive quantity $g$ with
preferred range $[g_{\rm lo},g_{\rm hi}]$, define
\begin{equation*}
\mathcal P(g)
=
\frac{\lambda}{2}
\left[
h\left\{\log\left(\frac{g_{\rm lo}}{g}\right)\right\}^2
+
h\left\{\log\left(\frac{g}{g_{\rm hi}}\right)\right\}^2
\right],
\qquad
h(x)=\max(x,0).
\end{equation*}
For quantities requiring only a lower threshold, only the first term
is used. These are soft penalties: values outside the preferred
region remain admissible.

The regularization is applied to driving-process variances, pole and
zero magnitudes, and AR decay rates.
Table~\ref{tab:regularization} summarizes the quantities regularized
during the first-stage optimization and the preferred regions induced
by the corresponding penalties. Stationarity, minimum phase, and
positive definiteness are enforced through parameterization rather
than penalization.

\begin{table}[t]
\centering
\caption{Quantities regularized during the first-stage optimization
and the preferred regions induced by the corresponding soft penalties.
The rate limits $\rho_{\min}$ and $\rho_{\max}$ are determined from
the observing design. Stationarity, minimum phase, and positive
definiteness of $V$ are enforced through the parameterization rather
than through these penalties.}
\label{tab:regularization}
\begin{tabular}{lll}
\hline
Quantity & Penalty & Preferred region \\
\hline
$\sigma_j^2$
    & two-sided
    & $[\sigma_{\rm lo}^2,\sigma_{\rm hi}^2]$ \\
$|\lambda_{r,j}|$
    & two-sided
    & $[\rho_{\min},\rho_{\max}]$ \\
$|z_{m,j}|$
    & two-sided
    & $[\rho_{\min},\rho_{\max}]$ \\
$\kappa_{r,j}$
    & lower threshold
    & $\geq \rho_{\min}$ \\
\hline
\end{tabular}
\end{table}

Let $T$ be the observational baseline and let $\Delta t$ be the
coarsest per-band median sampling interval. The design-informed rate
limits are
\[
\rho_{\min}=\frac{1}{c_{\rm span}T},
\qquad
\rho_{\max}=\frac{1}{c_{\rm cadence}\Delta t},
\]
with
\[
c_{\rm span}=0.2,
\qquad
c_{\rm cadence}=0.7.
\]
These are inverse-time reference rates, not Fourier frequencies and
not Rayleigh or Nyquist limits. In particular, the distinction from
the angular-frequency convention in the main text should be retained;
conversion between angular frequency and cycles per unit time uses
$f=\omega/(2\pi)$.

For an AR pole, the decay-rate penalty uses
\[
\mathcal P_\kappa
=
\frac{\lambda_\kappa}{2}
\sum_{j,r}
h\left\{
\log\left(\frac{\rho_{\min}}{\kappa_{r,j}}\right)
\right\}^2,
\qquad
\lambda_\kappa=10^{3}.
\]
Near pole--zero cancellation is not penalized at either stage. The
regularization therefore does not exclude or discourage such
configurations; they remain to be assessed by the ordinary likelihood
and model-selection criterion, after all first-stage penalties and the
covariance loading are removed.

During the first stage only, the driving covariance used in the
likelihood is loaded as
\[
V_\lambda
=
V+\lambda_V\frac{\operatorname{tr}(V)}{k}I_k,
\qquad
\lambda_V=0.05.
\]
Let $\ell_\lambda(\theta)$ be the resulting log-likelihood and
$\mathcal J(\theta)$ the sum of the applicable penalties. The
first-stage objective is
\[
Q(\theta)=\ell_\lambda(\theta)-\mathcal J(\theta).
\]
The loading and all penalties are removed before final inference.

\subsection{Optimization and convergence}
\label{sec:supp.optimization}

Both stages optimize in the unconstrained coordinates of
Section~\ref{sec:supp.parameterization}, so stationarity, minimum
phase, and positive definiteness hold at every point the optimizer
visits and no constraint is ever active at a solution. The stages
differ in their objective. The first maximizes the regularized,
loaded criterion $Q(\boldsymbol{\theta})$ of
Section~\ref{sec:supp.regularization}. The second maximizes the
ordinary, unloaded log-likelihood
$\ell(\boldsymbol{\theta})
=\log L(\boldsymbol{\theta};\mathbf{y}^{\ast})$.

\subsubsection{Starting values and the first-stage search}
\label{sec:supp.optimization.starts}

The first stage is run from several starting values for each
candidate order, and the solution with the best value of $Q$ is kept.
Four kinds of starting value are used.

The first is a band-wise warm start. Each band is fitted alone as a
univariate CARMA$(p,q)$ model, itself from four random starting
values, and the resulting per-band autoregressive coefficients,
moving-average coefficients, driving variances, and means are
assembled into a joint $\boldsymbol{\theta}$ whose off-diagonal
Cholesky entries are set to zero. The within-band dynamics are then
already near a univariate optimum when the joint optimization begins,
and only the cross-band structure has to be found from an
uninformative point. A band whose univariate fit fails is given a
negligible driving variance, and the warm start is abandoned if more
than $d/2$ bands fail.

The second is a set of perturbations of that warm start, Gaussian
with scale $10^{-3}$ in the unconstrained coordinates. These occupy
the first half of the remaining starting values and search the
immediate neighborhood of the band-wise solution.

The third is a set of draws from a sampler built on the observing
design. Autoregressive natural frequencies are drawn log-uniformly
over the range $[\rho_{\min},\rho_{\max}]$ of
Section~\ref{sec:supp.regularization}, damping ratios are drawn
uniformly over $[0.05,0.85]$ subject to the same lower threshold
$\zeta_j\omega_{n,j}\geq\rho_{\min}$ that the decay-rate penalty
uses, and moving-average zeros are placed relative to the
autoregressive frequencies of the same band so that the zero also
falls inside $[\rho_{\min},\rho_{\max}]$. Initialization and
regularization therefore use one set of reference rates rather than
two.

The fourth applies to $p\geq2$ only. One further starting value
copies the warm start but resets the autoregressive quadratic factor
of every band to an underdamped, low-frequency draw, with
$\zeta_j\in[0.10,0.40]$ and $\omega_{n,j}$ in the lower half of
$[\rho_{\min},\rho_{\max}]$. A local search does not cross between
damping regimes. Started from a nearly critically damped
autoregressive factor near the fast end of the resolvable range,
gradient-based optimization does not reach an underdamped
low-frequency pole, and an MCARMA$(2,1)$ fit started there can
terminate below the MCARMA$(2,0)$ fit with its moving-average
component collapsed. The extra starting value begins inside the
underdamped region so that the optimizer can descend from within it.

Each starting value is optimized by an unconstrained quasi-Newton
(BFGS) search. The gradient of $Q$ is computed analytically by
reverse-mode automatic differentiation through the Kalman recursion
rather than by finite differences, which would cost roughly
$\dim(\boldsymbol{\theta})+1$ filter passes per line-search step. The
iteration limit is 1000 and the gradient tolerance is $10^{-5}$. The
analyses reported here use six joint starting values per candidate
order, with four starting values inside each univariate fit of the
band-wise warm start.

\subsubsection{Second-stage refinement and its acceptance test}
\label{sec:supp.optimization.stage2}

The second stage uses one scoring function throughout: the ordinary
log-likelihood, with no penalty, no covariance loading, and, where a
deterministic per-band trend is estimated jointly rather than
subtracted beforehand, that same trend. Every quantity compared below
is computed by that function, so no comparison is made between two
different objectives.

To separate a numerical stopping point from a converged ordinary
maximum-likelihood solution, a candidate
$\widetilde{\boldsymbol{\theta}}$ is accepted only if
\begin{equation*}
\ell(\widetilde{\boldsymbol{\theta}})
\geq
\ell(\widehat{\boldsymbol{\theta}}_{\rm init})-\tau,
\qquad
\|\nabla\ell(\widetilde{\boldsymbol{\theta}})\|_\infty
\leq
\varepsilon
\max\{1,|\ell(\widetilde{\boldsymbol{\theta}})|\},
\end{equation*}
with $\tau=10^{-8}$ and $\varepsilon=10^{-4}$. The first condition
checks that ordinary-likelihood refinement has not numerically
decreased the objective below its own starting value, and the second
requires approximate stationarity. The stationarity threshold is
relative to the attained log-likelihood, so a fit with
$\ell\approx2000$ must reach a gradient sup-norm near $0.2$ rather
than near $10^{-4}$.

Stationarity is tested directly rather than read from the optimizer's
convergence flag. A quasi-Newton line search reports failure whenever
it terminates on loss of precision, which happens routinely at points
whose gradient sup-norm already satisfies the condition above.
Deciding on the flag would send almost every fit to the
derivative-free fallback for no reason. The flag is recorded but does
not determine acceptance.

Up to four attempts are made, in the order of
Table~\ref{tab:stage2.attempts}, and the first that satisfies both
conditions is accepted. Two mechanisms can let an attempt finish
below its own starting value, and the ordering removes them one at a
time. The automatically differentiated objective and the Kalman
evaluation of the same likelihood agree to six decimal places at
ordinary parameter values and disagree in the extremes, so descent in
one need not be descent in the other; the third attempt therefore
minimizes the Kalman evaluation itself, which is the function the
result is scored on. A multiple-start attempt can also return the
optimum of a different basin, because its starting values are ranked
by the objective that was minimized rather than by the scoring
function; the first attempt therefore ranks its starting values on
the scoring function, so that one function chooses the winner inside
the optimization and decides acceptance outside it. The last attempt
is a derivative-free simplex search, which retains its best vertex
and so cannot return a point below its start. It is weak in this many
dimensions, which is why it is last. Its solution is then polished by
a local quasi-Newton step, retained only when the polish does not
lower the score, both because a simplex method terminates on a small
simplex rather than a small gradient and because the curvature used
for standard errors is read off the quasi-Newton run.

All four attempts reparameterize the search by the curvature diagonal
at the starting point, computed once from the exact Hessian of the
scoring function. This changes the path the optimizer takes and
nothing else: the log-likelihood and gradient in the acceptance test
are evaluated in the original coordinates.

\begin{table}[b]
\centering
\caption{The second-stage attempts, in the order they are made, with
the number of converged SDSS order-fits each one produced. Counts are
over the $134\times3=402$ order-fits of
Table~\ref{tab:convergence.census} and sum to the 313 that converged.
Every attempt starts from the first-stage solution and is scored by
the same ordinary, unloaded log-likelihood.}
\label{tab:stage2.attempts}
\begin{tabular}{llr}
\hline\hline
Attempt & Search & Converged fits\\
\hline
1 & Multiple starting values, analytic gradient      & 206\\
2 & Single starting value, analytic gradient         & 8\\
3 & Single starting value, Kalman-evaluated objective & 86\\
4 & Derivative-free simplex, quasi-Newton polish     & 13\\
\hline
\multicolumn{2}{l}{Total converged} & 313\\
\hline
\end{tabular}
\end{table}

\subsubsection{Fits that do not converge}
\label{sec:supp.optimization.nonconvergence}

If no attempt satisfies both conditions, the first-stage solution is
returned, rescored under the ordinary unloaded likelihood, and
recorded as unconverged. It is not reported as a maximum likelihood
estimate. Three consequences follow, and they are handled separately.

\begin{table}[b]
\centering
\caption{Second-stage convergence over the 134 screened SDSS
Stripe~82 objects, each fitted at three candidate orders. The
selected-order count is the statistic usually quoted; the all-three
count is the one relevant to an AICc comparison, because a
non-stationary fit at a losing order biases that order's AICc upward.
The 62 objects converging at all three orders are a subset of the 99
converging at the selected order.}
\label{tab:convergence.census}
\begin{tabular}{lrr}
\hline\hline
Fits & Converged & Out of\\
\hline
MCARMA$(1,0)$              & 134 & 134\\
MCARMA$(2,0)$              &  89 & 134\\
MCARMA$(2,1)$              &  90 & 134\\
All order-fits             & 313 & 402\\
\hline
At the AICc-selected order &  99 & 134\\
At all three orders        &  62 & 134\\
\hline
\end{tabular}
\end{table}

First, the reported log-likelihood is still the ordinary unloaded
log-likelihood evaluated at the reported parameter value, so the same
function is used at every order and the values remain comparable
across orders. What has not been established is that the point is
stationary.

Second, the reported value is a lower bound on the maximum of that
order's ordinary likelihood. The AICc computed from it is therefore
too large, and the direction of the error is known: an unconverged
order is penalized in the comparison. Such an order can lose a
comparison it should have won, and it cannot win one it should have
lost.

Third, curvature at such a point is not used to report standard
errors. Near-singular or non-positive-definite observed information
is recorded as a failure of uncertainty quantification rather than
regularized to produce finite standard errors
(Section~\ref{sec:supp.uncertainty}).

The second consequence applies at every fitted order and not only at
the order AICc selects, so a convergence rate quoted at the selected
order understates how much of the comparison is exposed to it.
Table~\ref{tab:convergence.census} reports both for the 134 screened
Stripe~82 objects of Section~\ref{sec:sdss}. Of the
$134\times3=402$ order-fits, 313 converged. MCARMA$(1,0)$ converged
for every object, MCARMA$(2,0)$ for 89, and MCARMA$(2,1)$ for 90. The
selected order converged for 99 objects; all three orders converged
together for 62, a subset of those 99.

Of the 54 order-fits that did not converge at an order AICc did not
select, 44 are MCARMA$(2,0)$ and 10 are MCARMA$(2,1)$. The AICc
difference separating such a fit from the selected order has median
17.1 and minimum 0.46, and only one of the 54 lies within 2 of the
selected order. These differences are not small, but neither are they
beyond the reach of a second-stage gain:
Section~\ref{sec:supp.simulation.sensitivity} measures median gains
of 4.7 to 24.2 nats when an over-flexible candidate is fitted to data
from a simpler process, which is 9.4 to 48.4 on the AICc scale. The
direction of the bias can be established by argument and its size
cannot, so we measure the effect of the restriction directly.

Restricting the sample to the 62 objects that converged at all three
orders moves the distribution of selected orders in the predicted
direction and by a small amount. MCARMA$(1,0)$ falls from 18.7\% to
16.1\%, MCARMA$(2,0)$ is 3.0\% against 3.2\%, and MCARMA$(2,1)$ rises
from 78.4\% to 80.6\%, so support for dynamics beyond the first-order
model moves from 81.3\% to 83.9\%. Median pairwise coherence over the
restricted sample falls by 0.007, 0.014, 0.028, and 0.034 at band
separations of one through four, with the larger changes at the
larger separations where coherence is least well determined. The
three case studies of Section~\ref{sec:sdss} converged at all
three fitted orders, so no reported case-study result rests on an
unconverged fit.

\subsection{Additional uncertainty diagnostics}
\label{sec:supp.uncertainty}

For a derived quantity
$\boldsymbol{\psi}=g(\boldsymbol{\theta})$, the delta-method
covariance is
\[
\widehat{\operatorname{Var}}
(\widehat{\boldsymbol{\psi}})
=
J_g(\widehat{\boldsymbol{\theta}})
\widehat{\operatorname{Var}}
(\widehat{\boldsymbol{\theta}})
J_g(\widehat{\boldsymbol{\theta}})^\top.
\]
The observed information also provides an identifiability diagnostic:
near-singular curvature can arise when dynamical scales are poorly
resolved, poles and zeros nearly cancel, or covariance parameters
approach a boundary. Curvature calculations always use the ordinary
likelihood; the first-stage penalties and covariance loading are
excluded.

The simulation study evaluates empirical coverage of the
local Hessian-based intervals. Cases in which the information matrix
does not support reliable inversion are treated as uncertainty-
quantification failures rather than being silently regularized solely
to produce finite standard errors.

\subsection{Model-selection sensitivity}
\label{sec:supp.selection}

The main text uses AICc evaluated at the final ordinary MLE. We also
assess sensitivity to a generalized information criterion that
accounts for the effective dimension induced by the regularized
estimation stage. This sensitivity analysis is intentionally
secondary: the penalties are numerical devices rather than part of
the final statistical model.

We also monitor
\[
\Delta\ell
=
\ell(\widehat{\boldsymbol{\theta}})
-
\ell(\widehat{\boldsymbol{\theta}}_{\rm init}),
\]
with both terms evaluated under the ordinary, unloaded likelihood.
This separates sensitivity to initialization from sensitivity to the
model-selection rule. The numerical results and tables for the AICc
versus generalized-criterion comparison are reported with the
supplementary simulation results.

\section{Supplementary Results for the Simulation Study}
\label{sec:supp.controlled}

This appendix provides the detailed results supporting the simulation study in Section~\ref{sec:simulation}. The simulations generate five-band quasar time series under an
astronomy-motivated observing design that retains irregular sampling,
seasonal gaps, heteroscedastic measurement errors, and partially
observed bands, while the underlying stochastic dynamics and
measurement-noise level are varied systematically. This simulation setting complements the three real-data SDSS quasar case studies in the main text, for which the generative stochastic processes and parameters are unknown. The sections below provide complete definitions of the evaluation criteria, coefficient-wise and alternative spectral-recovery analyses, expanded weak-identifiability diagnostics, reconstruction between observed epochs, uncertainty calibration, sensitivity to numerical regularization and model-selection criterion, and computational benchmarks.

\subsection{Detailed simulation design and evaluation criteria}
\label{sec:supp.simulation.design}

We simulate five-band quasar time series from MCARMA$(1,0)$,
MCARMA$(2,0)$, and MCARMA$(2,1)$ processes. For each second-order model, we consider
underdamped, critically damped, and overdamped regimes in order to
evaluate inference under qualitatively different stochastic
dynamics. These regimes differ in the locations and types of the AR
poles and therefore generate substantially different temporal and
spectral behavior even at the same nominal model order.

Cross-band dependence is generated according to
\begin{equation*}
\rho_{j\ell}
=
0.9^{|j-\ell|},
\end{equation*}
so that neighboring bands are strongly correlated while dependence
decreases smoothly with band separation. The purpose of this
construction is not to reproduce a particular astronomical source
but to create a setting in which substantial cross-band information
is available to the joint multivariate fit.

All simulated data sets use the same astronomy-motivated observing
design, with irregular sampling, pronounced seasonal gaps, and one
band observed at each epoch. Observation times are drawn separately
for each data set rather than shared.
A data set contains 2439 to 2674 scalar observations, with a median
of 2527, or about 500 observations per band. The baseline is
bimodal: 61\% of the data sets span 1550 days and the remainder
extend to between 1832 and 1856 days. Quantities defined relative to
a data set's own baseline, such as the lower limit of the RISE
integration range, are therefore computed per data set. Because
only one band is observed at an epoch, the experiment also tests the
ability of the state-space likelihood to exploit dependence across
bands without requiring simultaneous observations.

Measurement uncertainties are varied to produce nominal variability
signal-to-noise ratios (S/N) of approximately 10, 4, and 2. We
define variability S/N as the stationary standard deviation of the
intrinsic stochastic process divided by the per-epoch measurement
uncertainty. This quantity measures the strength of the stochastic
variability relative to the observational noise and is distinct from
a flux-based detection S/N. The driving covariance is rescaled after
specification of the temporal dynamics so that these nominal
variability S/N levels are comparable across the different
generative settings rather than changing inadvertently with model
order or damping regime.

For each combination of generative order, damping regime, and
variability S/N, we generate 33 independent data sets. For each
second-order model there are therefore three damping regimes and
three variability S/N levels. Because MCARMA$(1,0)$ has no damping
classification, its three nominal damping cells at a given
variability S/N carry the same 33 generative processes, each observed
through an independent realization, so that its numerical summaries
can be compared on the same layout as those of the second-order
models. The 297 first-order data sets therefore rest on 33 distinct
parameter vectors rather than 297, and dispersion within a
first-order cell reflects realization variability at a repeated set
of generative values. Across variability S/N levels the generative
parameters are held fixed for every order, so comparisons between S/N
levels are paired on the parameter values. The complete simulation
corpus contains 891 data sets.

For every data set, we fit MCARMA$(1,0)$, MCARMA$(2,0)$, and
MCARMA$(2,1)$ using the two-stage estimation procedure of
Section~\ref{sec:estimation}, and select the preferred order using
AICc computed from the final ordinary likelihood. To assess
estimation accuracy independently of model selection, we also
evaluate fits at the generative order. At that order, the
corresponding univariate CARMA models are fitted separately to each
band. Because each marginal PSD has the same CARMA functional form
under the multivariate and univariate models, this comparison
isolates the gain from jointly estimating correlated bands rather
than a difference in the marginal stochastic model.

We evaluate model-order recovery by the proportion of simulated data
sets for which AICc selects the generative order. Parameter accuracy
is evaluated using relative estimation error. For a scalar parameter
$\theta$ with generative value $\theta_0$, we define
\begin{equation*}
E_{\theta}
=
\frac{|\widehat{\theta}-\theta_0|}
     {|\theta_0|}.
\end{equation*}
We use this measure for individual model coefficients as well as
derived quantities such as characteristic frequencies and
timescales. In particular, the comparisons below use
$\alpha_{1,1}$ as a representative marginal dynamic parameter when
contrasting multivariate and univariate estimation.

Individual AR and MA parameters can be weakly identified even when
their combined effect on the stochastic process is much better
determined. We therefore place particular emphasis on recovery of
the marginal power spectral densities. Let $P_{jj}(f)$ and
$\widehat P_{jj}(f)$ denote the generative and estimated
marginal PSDs for band $j$. Both measures below are evaluated on
frequency grids in cycles per day, so the argument here is
$f=\omega/(2\pi)$ rather than the angular $\omega$ of
Section~\ref{sec2:spectral}. Both are ratios of integrals of the same
spectra, so neither depends on the normalization convention. We use
two complementary measures of
spectral accuracy, one emphasizing spectral shape and the other
retaining both shape and normalization.

The shape-normalized spectral error (SNSE) is defined in Section~\ref{sec:simulation.design}. It normalizes each marginal spectrum by its value at the lowest evaluated frequency, uses the common range $10^{-3.5}$ to $10^{-0.3}$ cycles day$^{-1}$, and is averaged over all five bands. We therefore use SNSE as the primary measure of spectral-shape recovery.

Our second measure, the relative integrated squared error (RISE),
compares the estimated and generative marginal spectra on the
logarithmic scale without normalizing away their overall level:
\begin{equation*}
\operatorname{RISE}_j
=
\frac{
\displaystyle
\int_{f_{\rm lo}}^{f_{\rm hi}}
\left[
\log \widehat P_{jj}(f)
-
\log P_{jj}(f)
\right]^2\,df
}{
\displaystyle
\int_{f_{\rm lo}}^{f_{\rm hi}}
\left[
\log P_{jj}(f)
\right]^2\,df
}.
\end{equation*}
For the simulations,
\begin{equation*}
f_{\rm lo}=\frac{1}{T},
\qquad
f_{\rm hi}=0.15
\end{equation*}
cycles day$^{-1}$, where $T$ is the observational baseline. RISE is
averaged over all five bands. Unlike SNSE, it retains discrepancies
in both the overall spectral level and spectral shape and is
evaluated over a range tied to the temporal extent of the data.

The two spectral measures therefore answer different questions.
Both average over all five bands, but
SNSE uses a common frequency range and removes spectral
normalization, making it useful for comparing recovery of spectral
shape across simulation settings. RISE retains level information and
uses an observationally motivated frequency range, making it more
directly sensitive to recovery of the full marginal stochastic
spectrum. We use both because accurate recovery of an individual
coefficient does not necessarily imply accurate recovery of the
spectrum, and conversely the spectrum can remain well estimated over
the observationally informative range even when individual
coefficients are weakly identified.

SNSE is a linear, shape-only comparison on a fixed grid; RISE is a
logarithmic, level-sensitive comparison on a per-data-set range. The
two are therefore not on a common scale and typically differ by about
two orders of magnitude on the same fit. We report them separately and
carry no quantity across them.

We additionally evaluate weak identifiability through characteristic
timescales and pole--zero separation, uncertainty calibration through
coverage and standardized estimation errors, sensitivity to the
first-stage regularization and subsequent ordinary-likelihood
refinement, sensitivity to the information criterion, and
computational cost. The corresponding definitions and results are
introduced with each experiment below.

\subsection{Additional parameter and spectral recovery results}
\label{sec:supp.simulation.multivariate}

We next evaluate estimation accuracy at the generative model order and
ask whether jointly modeling the five correlated bands improves
inference relative to fitting each band separately. Because each
band has the same marginal CARMA$(p,q)$ form under the multivariate
and univariate analyses, differences in accuracy arise from the use
of cross-band dependence rather than from differences in the marginal
model specification. The comparison therefore isolates the extent to
which observations in correlated bands provide information about one
another's latent stochastic dynamics.

\begin{figure}[t]
\centering
\includegraphics[width=\textwidth]{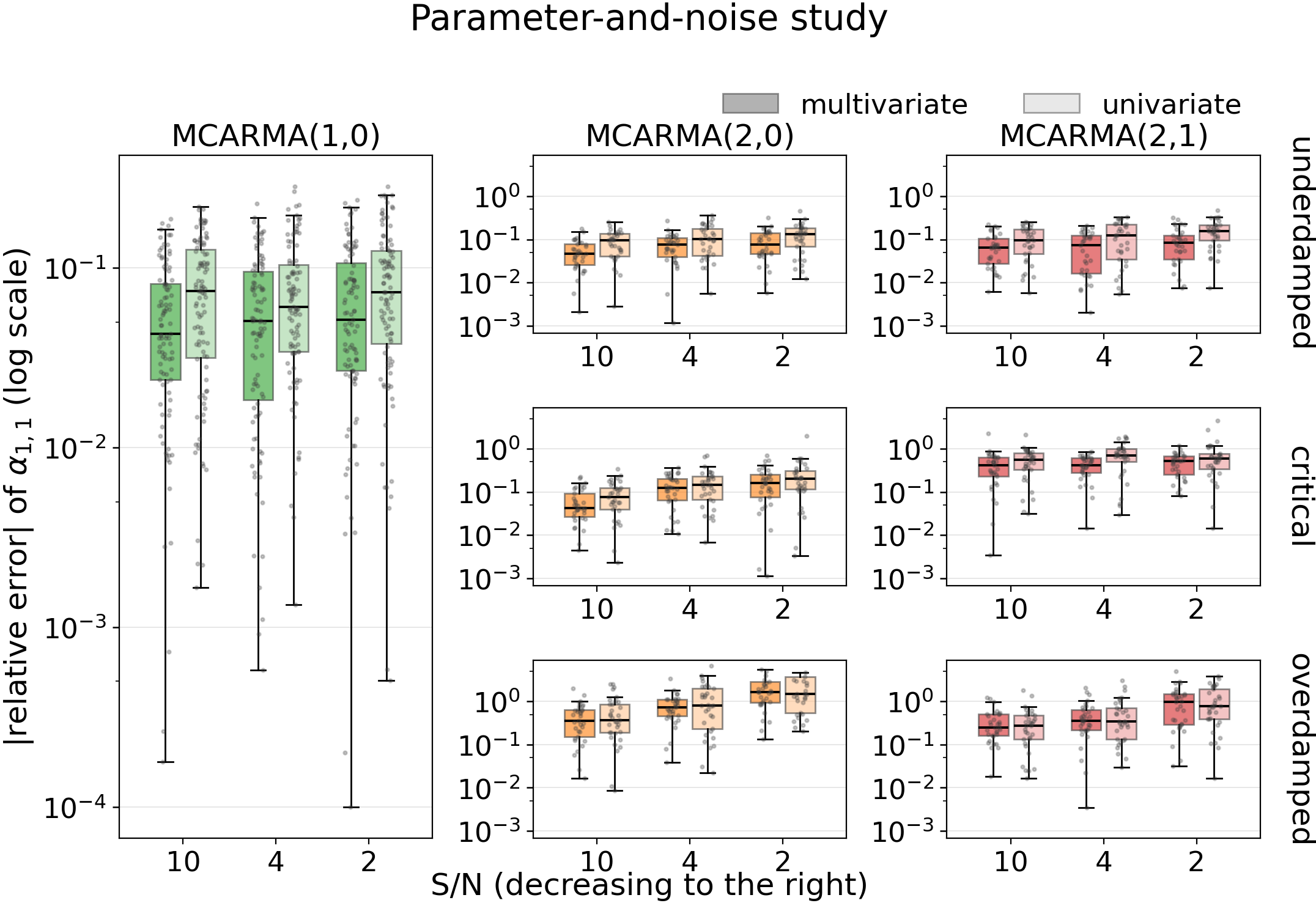}
\caption{Parameter-estimation accuracy in the simulation
study. Relative error in $\alpha_{1,1}$ is compared between the joint
multivariate MCARMA fit and separate univariate CARMA fits at the
generative order across the model-order, damping-regime, and
variability-S/N settings. The MCARMA$(1,0)$ panels use the
corresponding first-order realizations across the common simulation
layout, whereas each second-order cell contains independently
generated data sets for the indicated damping regime. Boxes show
quartiles, whiskers extend to 1.5 times the interquartile range, and
individual data sets are shown as points. Smaller values indicate
more accurate estimation.}
\label{fig:param_accuracy}
\end{figure}

We first consider recovery of an individual dynamical parameter.
Figure~\ref{fig:param_accuracy} compares the relative error in
$\alpha_{1,1}$ under the multivariate and separate univariate fits
across the 27 combinations of generative order, damping regime, and
variability S/N. The multivariate fit has smaller error in 23 of the
27 settings, with a median ratio of univariate to multivariate error
of approximately 1.3. Thus, information sharing across correlated
bands generally improves estimation of the marginal dynamics,
although the improvement for an individual coefficient is moderate.

The improvement is more pronounced when estimation is evaluated
through the implied stochastic spectrum. The SNSE comparison in Section~\ref{sec:simulation.multivariate} compares the shape-normalized
spectral error under the multivariate and univariate fits.
The multivariate fit has smaller median SNSE in 26 of the 27
simulation settings, and the median ratio of univariate to
multivariate SNSE across the settings is approximately 1.9. The
advantage occurs at all three variability-S/N levels rather than
being confined to the noisiest data sets. Thus, the gain from
cross-band information is not simply a noise-reduction effect that
vanishes when the individual bands are well measured.

In several higher-order settings, separate univariate fits can also
produce spectral features that are reduced or absent under joint
estimation. Because the multivariate model is constrained
simultaneously by several correlated realizations of the underlying
variability, poorly supported features in one band can be moderated
by information from the remaining bands. This effect is especially
relevant for higher-order models, in which modest errors in several
AR and MA parameters can combine to produce appreciable changes in
the implied PSD.

The same conclusion is obtained using RISE, which retains differences
in both spectral level and shape. Table~\ref{tab:psd_rise} aggregates
the comparison over individual data sets rather than first
summarizing within the 27 design cells. For every generative order,
the mean RISE is smaller under the multivariate fit. The multivariate
fit also has smaller RISE for 217 of 297 MCARMA$(1,0)$ data sets,
244 of 297 MCARMA$(2,0)$ data sets, and 244 of 297 MCARMA$(2,1)$
data sets. The corpus contains 297 data sets of each generative
order, and every one of them enters the comparison.

\begin{table}[t]
\centering
\caption{Mean relative integrated squared error (RISE) of the
estimated marginal PSDs in the simulation study. The
last column gives the number of data sets for which the joint
multivariate MCARMA fit has smaller RISE than the corresponding
separate univariate CARMA fits. Smaller values indicate more accurate
spectral recovery. Each generative order contributes 297 data sets to
the corpus.}
\label{tab:psd_rise}
\begin{tabular}{@{}lccc@{}}
\hline\hline
& \multicolumn{2}{c}{Mean RISE} & \\
\cline{2-3}
Generative order & MCARMA & Univariate & MCARMA smaller\\
\hline
$(1,0)$ & 0.0037 & 0.0052 & 217/297\\
$(2,0)$ & 0.0064 & 0.0167 & 244/297\\
$(2,1)$ & 0.0068 & 0.0133 & 244/297\\
\hline
\end{tabular}
\end{table}

\begin{figure}[b]
\centering
\includegraphics[width=\textwidth]{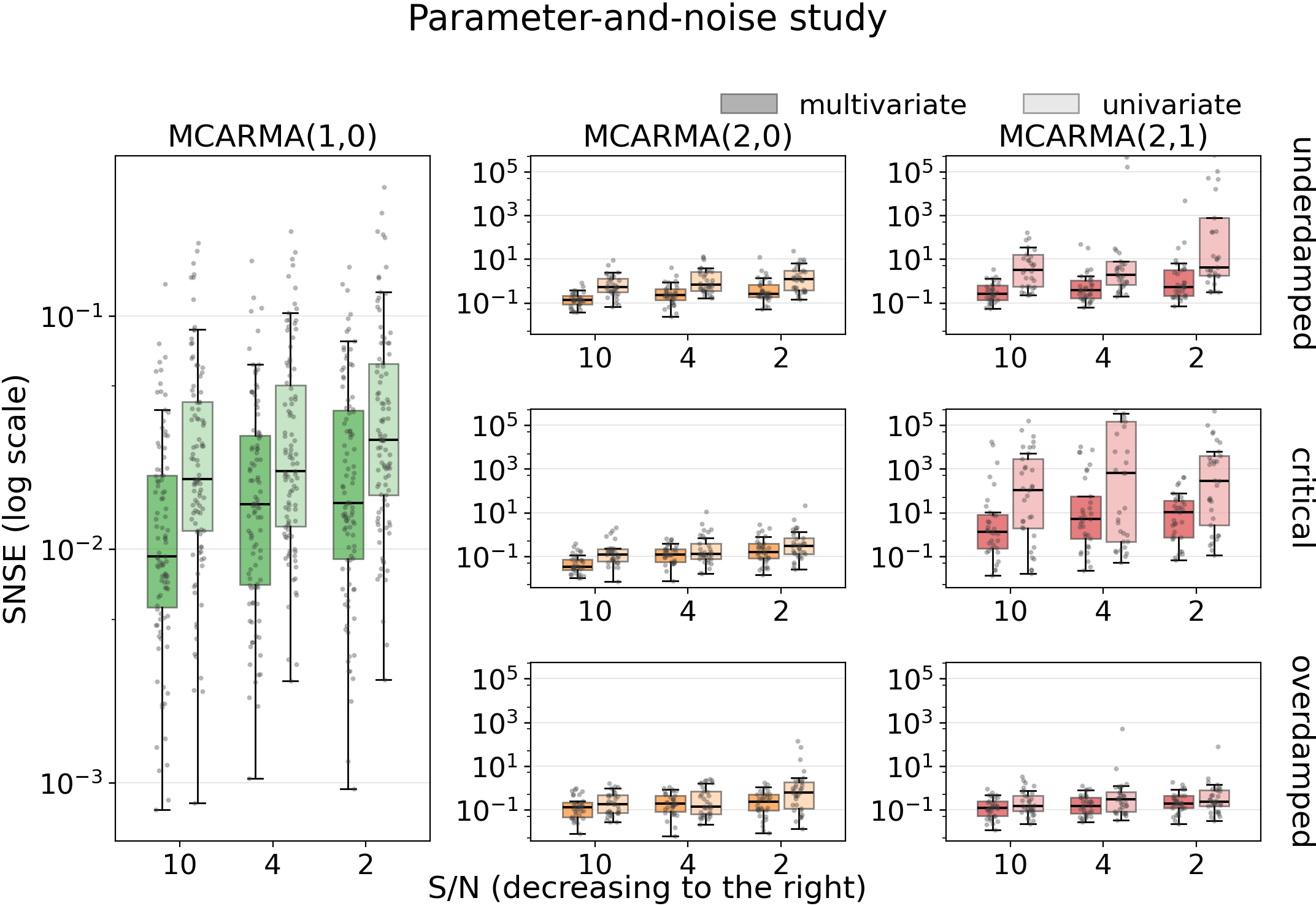}
\caption{Distribution of shape-normalized spectral error (SNSE) over
the parameter-and-noise simulation corpus, comparing
joint multivariate MCARMA and separate univariate CARMA fits. SNSE
is averaged over all five bands and displayed using the
same model-order, damping-regime, and variability-S/N layout as
Figure~\ref{fig:param_accuracy}. The vertical axis is logarithmic
because SNSE spans approximately four orders of magnitude. Taking
the median within each of the 27 simulation settings, the
multivariate fit has smaller SNSE in 26 settings, with a median
univariate-to-multivariate ratio of approximately 1.9.}
\label{fig:corpus.rise}
\end{figure}

Figure~\ref{fig:corpus.rise} provides a complementary view of the
full distribution of SNSE over the parameter-and-noise simulation
corpus. The spectral errors span approximately four orders of
magnitude, demonstrating that the absolute difficulty of spectral
estimation varies substantially across model orders, damping regimes,
and measurement-noise levels. A logarithmic vertical scale is
therefore required; on a linear scale, most of the distributions
would be compressed near zero.

The cell-by-cell SNSE comparison and the data-set-level RISE
comparison therefore give the same qualitative conclusion. The
multivariate advantage is broadly distributed across the simulation
corpus rather than being driven by a small number of especially
favorable settings. In particular, all three generative orders have
smaller mean RISE under joint fitting, while the shape-based SNSE
comparison shows improvement in nearly every combination of
dynamical regime and variability S/N.

The larger gain for spectral recovery than for a single coefficient
has a natural interpretation. The marginal PSD is determined jointly
by the full set of AR and MA parameters and the stochastic driving
amplitude. Modest improvements in several estimated dynamical
quantities can therefore combine to produce a substantially more
accurate estimate of the overall stochastic variability. Conversely,
an individual coefficient can be relatively poorly determined while
different parameter combinations still imply similar spectra over
the frequencies directly informed by the observations.

These results demonstrate that the benefit of the multivariate model
is not restricted to estimation of cross-band dependence. Although
the marginal CARMA model of each band is unchanged, observations in
correlated bands provide additional information about its latent
stochastic evolution. Joint estimation therefore improves inference
for the marginal dynamics themselves, with the clearest improvement
appearing in recovery of the implied spectrum.

\subsection{Weak identifiability and temporal resolution}
\label{sec:supp.simulation.identifiability}

The preceding experiments reveal two conceptually distinct sources of
weak identification. First, a characteristic stochastic timescale may
lie near or beyond the temporal range effectively resolved by the
observing design. This is an observational limitation that can arise
even for a correctly specified low-order model. Second, a higher-order
model can be intrinsically close to a lower-order representation
through pole--zero cancellation. We examine these two mechanisms
separately.

To isolate the first mechanism, we conduct an extended
MCARMA$(1,0)$ experiment in which the model order and observing design
are held fixed while the characteristic timescale is varied. We
consider nominal characteristic timescales of 10, 30, 100, and
300 days, progressively moving the stochastic break toward the
low-frequency boundary informed by the observations. Because
MCARMA$(1,0)$ contains only one AR pole and no MA zero, deterioration
in this experiment cannot be attributed to pole--zero cancellation.

The nominal timescales define the centers of the corresponding
generative ranges rather than fixing every simulated process to an
identical value. Each timescale setting contains 99 independent
generative processes, consisting of 33 realizations at each of the
three variability-S/N levels. The realizations are paired across
timescale settings by seed, so that corresponding simulations use
the same relative position within the generative parameter range and
the same observing design. This construction makes the comparison
across timescales less sensitive to unrelated differences among
simulated realizations.

The extended-timescale table in Section~\ref{sec:simulation.order} summarizes correct-order recovery,
median relative error in $\alpha_{1,1}$, median spectral error, and
the fraction of simulations classified as stationary in every band
by the stationarity diagnostic used in the real-data screening. The
first three settings lie within the primary operating range of the
simulation design. The 300-day setting straddles the practical
long-timescale boundary, approximately
$c_{\rm span}T=310$ days for a 1550-day baseline, and is therefore
included as a boundary case rather than as an ordinary operating
point.

Correct-order recovery decreases from 0.919 to 0.909 and 0.818
across the first three timescale settings and to 0.667 at the
300-day boundary case. Thus, even as the characteristic timescale
increases by approximately an order of magnitude within the primary
range, a slow MCARMA$(1,0)$ process is still selected as first order
in most simulations. The relatively high recovery rate of the
first-order model in the model-order recovery table in Section~\ref{sec:simulation.order} is therefore not
simply an artifact of considering a narrow range of easily resolved
first-order timescales.

\begin{figure}[b]
\centering
\includegraphics[width=0.49\textwidth]
{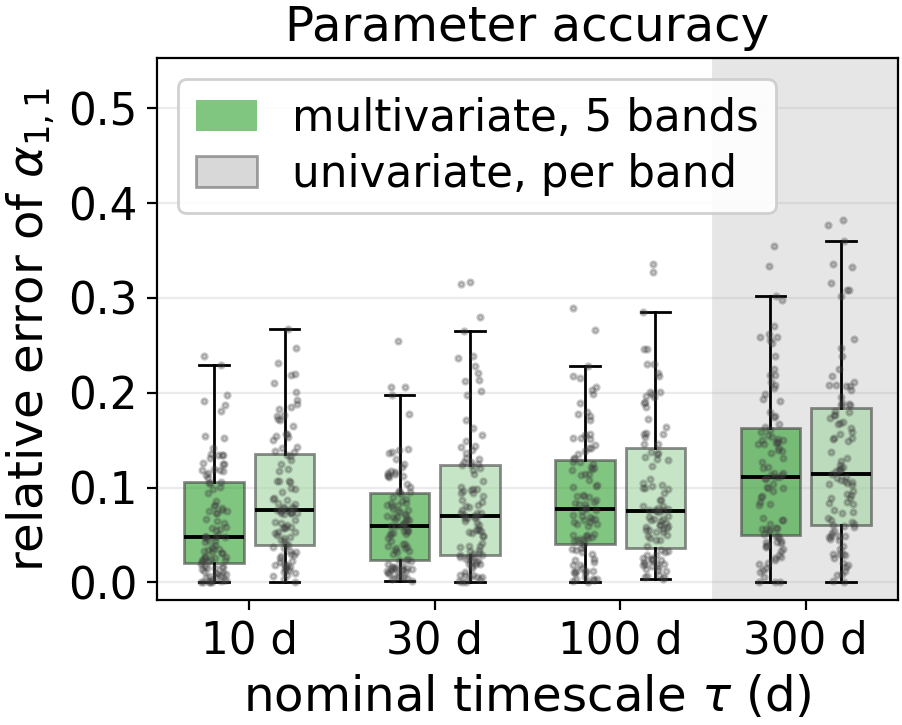}
\includegraphics[width=0.49\textwidth]
{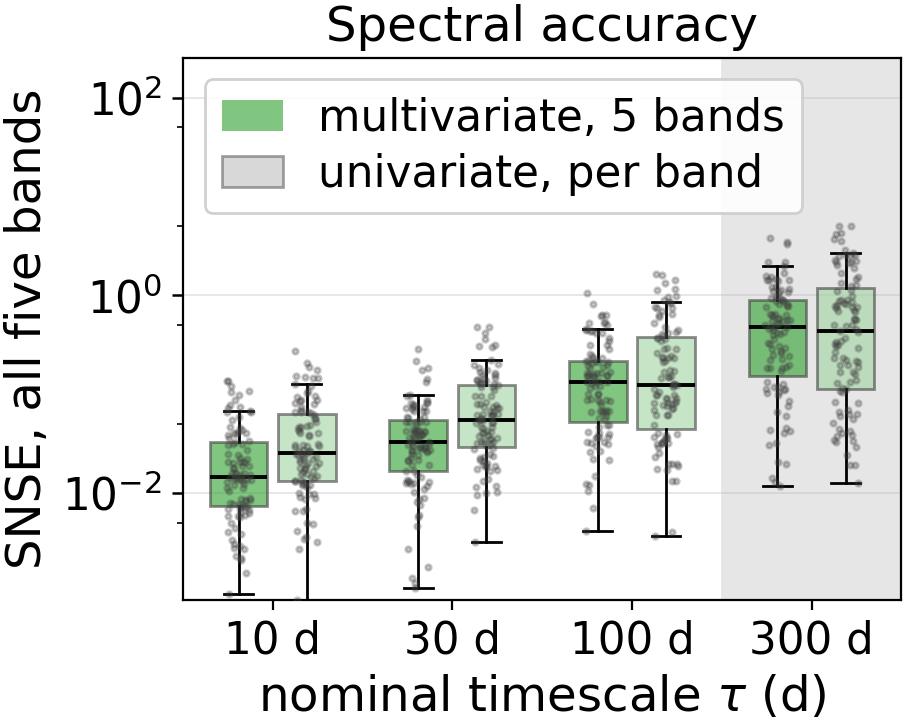}
\caption{Accuracy in the extended MCARMA$(1,0)$ timescale experiment.
Left: relative error of $\alpha_{1,1}$. Right: shape-normalized
spectral error (SNSE) on a logarithmic scale. Joint multivariate
MCARMA and separate univariate CARMA fits are compared over nominal
characteristic timescales of 10, 30, 100, and 300 days. The
300-day setting straddles the practical long-timescale boundary
$c_{\rm span}T=310$ days for this observing baseline and is treated
as a boundary case. Boxes show quartiles, whiskers extend to 1.5
times the interquartile range, and individual data sets are shown as
points.}
\label{fig:extended_timescales}
\end{figure}

Parameter and spectral recovery nevertheless deteriorate much more
substantially. Over the first three timescale settings, the median
relative error in $\alpha_{1,1}$ increases from 0.048 to 0.077,
whereas the median spectral error rises from approximately 0.015 to
0.131. Thus, spectral recovery degrades much faster than recovery of
the single AR coefficient as the characteristic timescale moves
toward the long-timescale boundary. The dense sampling of the
simulation design, with approximately 500 observations per band,
continues to constrain the local AR behavior reasonably well, while
the finite baseline provides increasingly little information
about the low-frequency turnover of the PSD. Figure~\ref{fig:extended_timescales} shows the corresponding
distributions of parameter and spectral error.

The multivariate advantage also diminishes as the characteristic
timescale increases. At the shortest setting, the median relative
error in $\alpha_{1,1}$ is approximately 0.048 under the
multivariate fit compared with 0.076 under the corresponding
univariate fits. By the 100-day setting the errors are essentially
the same, approximately 0.077 and 0.075, respectively. This behavior
has a direct information-sharing interpretation: when no individual
band contains substantial information about variability below the
low-frequency range resolved by the baseline, the remaining bands
have little additional long-timescale information to contribute.
Cross-band dependence can share information that is present in other
bands, but it cannot create information about temporal scales that
are unresolved in all of them.

The stationarity diagnostic responds even more sharply to increasing
timescale. The fraction of simulated data sets classified as
stationary in every band decreases from 1.00 to 0.78, 0.18, and
0.02 over the four settings. All of the simulated processes are
stationary by construction, so this decline reflects the inability
of a finite observing window to provide convincing evidence for
stationarity when the characteristic timescale becomes long relative
to the baseline; it is not a change in the underlying stochastic
process. By the nominal 100-day setting, the diagnostic fails to
classify 82\% of the realizations as stationary in every band.
Because the same type of screening is used before fitting the
real-data examples, this experiment also shows that a stationarity
screen can implicitly select against sufficiently slowly varying
processes.

The extended experiment therefore emphasizes that model-order
recovery, parameter estimation, spectral recovery, and empirical
stationarity diagnostics answer different questions and can
deteriorate at substantially different rates. A data set may contain
sufficient information to prefer MCARMA$(1,0)$ over a higher-order
alternative while providing much less information about the precise
location of its characteristic timescale or the low-frequency form
of its PSD. Correct model-order selection should consequently not be
interpreted as evidence that every characteristic timescale is
precisely estimated.

The second source of weak identification is structural rather than
primarily observational. For MCARMA$(2,1)$, define the nearest
pole--zero separation as in the definition in Section~\ref{sec:simulation.order}. As an MA zero approaches an AR
pole, the corresponding factors increasingly cancel in
\[
\frac{M_j(z)}{A_j(z)},
\]
and the process approaches an effectively lower-order representation.
In the limiting case of exact cancellation, the common factor drops
out of the transfer function and the nominal MCARMA$(2,1)$
representation is nonminimal. Because the MA zero in the formulation
considered here is real, exact cancellation can occur with a real AR
pole; a genuinely complex-conjugate AR pair cannot have only one
member exactly cancelled while retaining real-valued model
coefficients.

The MCARMA$(2,1)$ simulation corpus provides direct empirical
evidence for this mechanism. As reported in
Section~\ref{sec:simulation.order}, correct-order recovery is only
0.44 in the lowest quartile of pole--zero separation, increasing to
0.81, 0.95, and 0.95 in the successive quartiles. The median natural
frequency changes little across these quartiles. Thus, the sharp
difference in recoverability cannot be explained simply by movement
of the characteristic frequency toward the observational boundary;
it is specifically associated with proximity of an AR pole and MA
zero.

The distinction between these two mechanisms is important.
Observational weak identification arises because the available
baseline, cadence, or measurement precision provides insufficient
information about a dynamical scale. In principle, it can be reduced
by a more informative observing design. Pole--zero near-cancellation,
by contrast, reflects the geometry of the stochastic model itself.
Even highly informative observations can provide little information
for separately estimating two factors whose effects on the transfer
function nearly cancel. Near cancellation is therefore a structural
source of weak identifiability rather than merely another consequence
of sparse or noisy observations.

Both mechanisms also explain why parameter recovery need not coincide
with recovery of the stochastic process. Near an observational
boundary or a cancellation configuration, likelihood curvature in
individual parameter directions can become weak while combinations
of those parameters still determine the covariance or PSD reasonably
well over the frequencies informed by the observations. This is one
reason that the spectral criteria in
Section~\ref{sec:simulation.multivariate} provide an important
complement to coefficient-wise error.

Taken together, these experiments indicate that weak identification
should not be diagnosed solely from nominal model order or optimizer
convergence. The locations of the fitted poles and zeros relative to
one another and to the temporal range informed by the observations
provide additional information about which fitted dynamical features
can be interpreted reliably. This issue will also be reflected in
the likelihood-curvature diagnostics and uncertainty calibration
considered next.

\subsection{Reconstruction between observed epochs}
\label{sec:supp.simulation.rts}

The preceding experiments evaluate recovery of model order,
parameters, and spectral structure. A complementary question is how
the fitted stochastic process interpolates between observed epochs,
particularly across gaps in the observing sequence. We examine this
behavior using the Rauch--Tung--Striebel (RTS) smoother, which
computes the conditional distribution of the latent state given the
entire observed time series.

\begin{figure}[b]
\centering
\includegraphics[width=\textwidth]{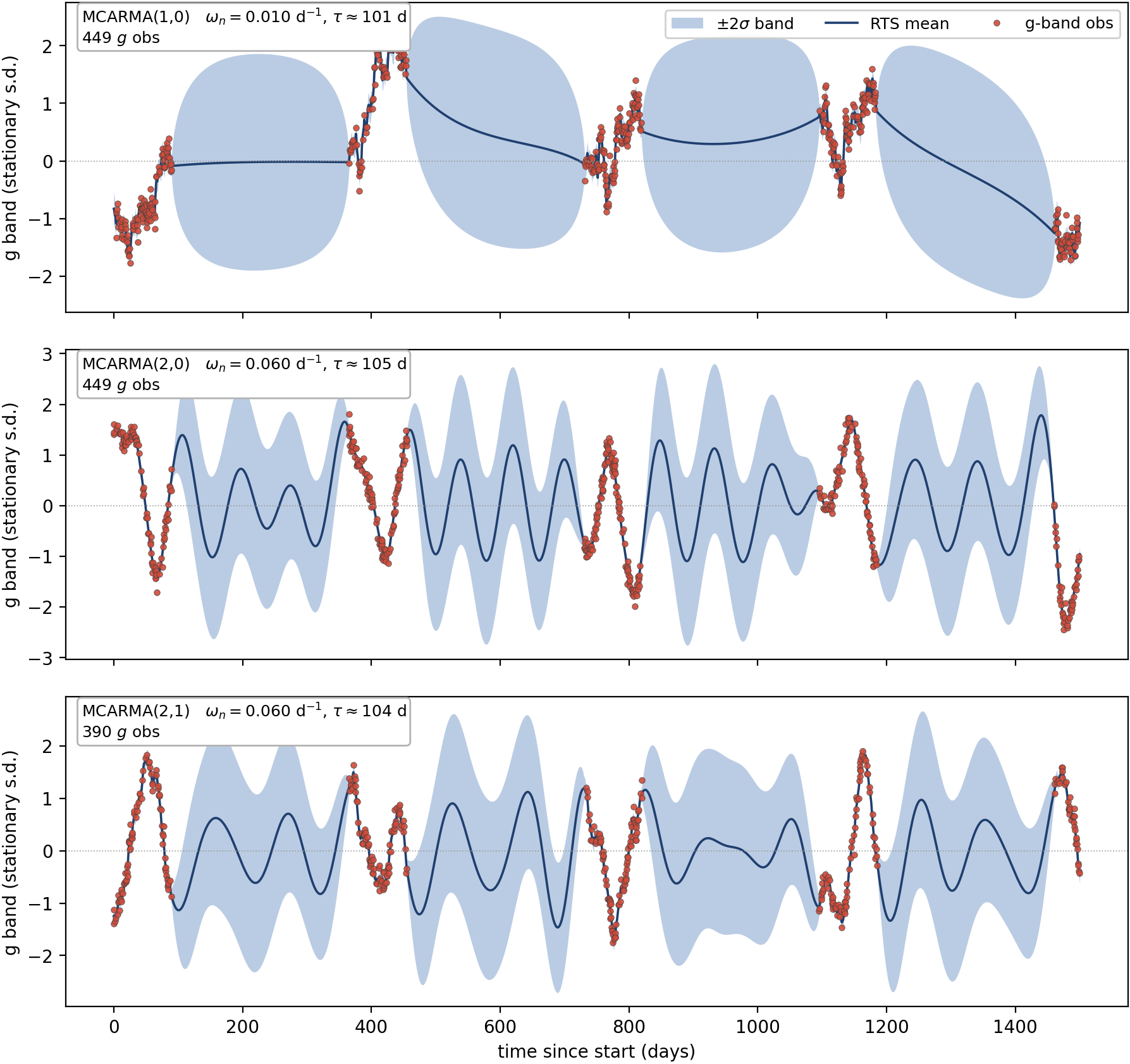}
\small
\caption{Reconstruction between observed epochs using the
Rauch--Tung--Striebel (RTS) smoother in the simulation
study. The smoother is evaluated at the generative parameters,
isolating differences in reconstruction implied by the stochastic
models from parameter-estimation error. From top to bottom, the
panels show MCARMA$(1,0)$, MCARMA$(2,0)$, and MCARMA$(2,1)$
processes under the parameter-and-noise simulation design.
Points denote observed $g$-band measurements, solid curves show the
RTS conditional means, and shaded regions give the corresponding
$\pm2\sigma$ conditional bands. The smoother is multivariate, so each
panel is reconstructed from observations in all five bands and not
only from the plotted one. Each panel is standardized by its own
stationary standard deviation, so a band approaching $\pm2$
corresponds to the unconditional stationary uncertainty.}
\label{fig:rts}
\end{figure}

To separate the intrinsic reconstructability of the stochastic
process from uncertainty introduced by parameter estimation, we
evaluate the smoother at the generative parameter values. The
resulting conditional means and variances therefore characterize what
can be inferred between observations when the stochastic model itself
is known. Differences among the reconstructed trajectories can then
be attributed to the dynamical structure of the MCARMA process and
the observing pattern rather than to estimation error.

Figure~\ref{fig:rts} shows representative
reconstructions for the three model orders. Within densely observed
parts of the time series, all three processes are reconstructed
similarly and the conditional uncertainty remains small. The
differences become more pronounced across longer observational gaps,
where the conditional evolution is increasingly determined by the
dynamics implied by the fitted poles and zeros.

For MCARMA$(1,0)$, the conditional mean relaxes monotonically toward
the stationary mean as the time since the last informative
observation increases. This behavior follows from the single
real-valued decay scale of the first-order process. In contrast,
an underdamped second-order process can retain oscillatory structure
across a gap through its complex-conjugate AR poles. The conditional
mean may therefore continue to exhibit coherent variation for a
longer interval even in the absence of direct observations.

The MCARMA$(2,0)$ and MCARMA$(2,1)$ processes also exhibit
noticeably different reconstructions across the observational gaps.
Although both models share second-order autoregressive dynamics, the
additional MA component in MCARMA$(2,1)$ modifies the covariance
structure and consequently changes both the conditional trajectory
and its uncertainty. The MA zero is most directly interpreted through
its effect on the spectral response, but this comparison illustrates
that its influence can also appear in time-domain conditional
prediction. Because the smoother is evaluated at the generative
parameters, however, these visible differences demonstrate
model-implied reconstruction behavior rather than the ability to
distinguish the two model orders visually when their parameters are
unknown.

The smoother uncertainty also provides a direct measure of how much
information the observing pattern supplies between observations.
Near observed epochs, conditioning sharply reduces uncertainty
relative to the stationary variance. As a gap becomes longer, the
conditional covariance gradually approaches the unconditional
stationary covariance because the process increasingly loses memory
of the surrounding observations. The rate and form of this return
depend on the characteristic dynamics of the model.

This experiment therefore complements the parameter- and frequency-domain diagnostics above. The spectral analysis characterizes which stochastic features are supported globally by the data, whereas the RTS smoother shows how those same dynamics determine local reconstruction in time. Together they illustrate why higher-order stochastic structure is most reliably assessed through the likelihood and spectral representation rather than solely by visual comparison of latent-process reconstructions.

\subsection{Uncertainty calibration}
\label{sec:supp.simulation.uncertainty}

We next evaluate the local likelihood-based uncertainty estimates of
Section~\ref{sec:uncertainty}. All calibration calculations are
performed at the generative model order so that lack of coverage is
not confounded with deliberate model misspecification. A
misspecified candidate model has no reason to provide intervals that
cover parameters of a different generative process, so only
generative-order fits are relevant for this calibration experiment.

The intentionally near-cancelling MCARMA$(2,1)$ setting is excluded
from the aggregate summaries because it is constructed specifically
to exhibit weak structural identifiability rather than ordinary
finite-sample calibration behavior. Including this setting would
therefore mix a known identification failure with the question of
whether the local likelihood approximation is appropriately
calibrated under otherwise identifiable generative settings.
Calibration statistics are computed over fits returning finite
standard errors.

For a scalar parameter $\theta$ with generative value $\theta_0$, we
consider the empirical coverage of the nominal $2\sigma$ interval and
the squared standardized error
\begin{equation*}
z^2
=
\left[
\frac{\widehat{\theta}-\theta_0}
     {\widehat{\operatorname{se}}(\widehat{\theta})}
\right]^2.
\end{equation*}
Under an approximately Gaussian and well-calibrated local likelihood
approximation, the nominal $2\sigma$ interval has coverage close to
0.95 and $z^2$ approximately follows a $\chi^2_1$ distribution, whose
median is approximately 0.455. A median $z^2$ substantially above
0.455 indicates that the estimation error is large relative to the
reported standard error. This may arise because the interval is too
narrow, because the point estimate is displaced from the generative
value, or because both effects occur simultaneously. A median
$z^2$ below 0.455 is consistent with standardized errors that are
smaller than expected under the reference Gaussian approximation.

The standard errors used here are based on the analytic Hessian of
the ordinary log-likelihood evaluated at the final
ordinary-likelihood estimate, rather than on an inverse-Hessian
approximation returned internally by the optimizer. This distinction
is important because the optimizer approximation is constructed for
numerical optimization rather than uncertainty quantification and
need not reproduce the full local dependence among model
parameters. In particular, inference for quantities such as the
cross-band correlation requires the relevant off-diagonal curvature
information.

Two details of this calculation should be stated. The Hessian is
evaluated after a short additional optimization from the reported
estimate to a stationarity tolerance tighter than the one used to
accept it, so the point of evaluation is a refinement of the reported
estimate and not literally that estimate. And when the observed
information is not positive definite, the simulation study forms the
covariance matrix by pseudoinversion, so a standard error is returned
in cases where the curvature is singular in some direction. The
fraction of fits with a positive-definite observed information is
therefore reported alongside every calibration statistic, and standard
errors from the remaining fits should be read with that in mind.

Table~\ref{tab:controlled.calibration} summarizes the calibration
results for the simulations. We report results for the
first-band AR coefficient $\alpha_{1,1}$, the adjacent-band
correlation $\rho_{12}$, and, for MCARMA$(2,1)$, the first-band MA
coefficient $b_{1,1}$. The table also gives the fraction of fitted
solutions for which the observed information is positive definite.

\begin{table}[b]
\centering
\caption{Uncertainty calibration at the generative model order in the
simulation study. ``mv'' denotes the joint multivariate
MCARMA fit and ``univ'' the corresponding single-band CARMA fit
evaluated on band 1 of the same data set. Reported diagnostics are
empirical $2\sigma$ coverage, median squared standardized error
$z^2$, and the fraction of fits with positive-definite local
curvature. A well-calibrated Gaussian local approximation has
approximately 0.95 coverage and median $z^2=0.455$. The
intentionally near-cancelling MCARMA$(2,1)$ setting is excluded.}
\label{tab:controlled.calibration}
\footnotesize
\setlength{\tabcolsep}{4pt}
\begin{tabular}{llccc}
\hline\hline
Measure & Fit & $(1,0)$ & $(2,0)$ & $(2,1)$\\
\hline
$2\sigma$ coverage of $\alpha_{1,1}$
    & mv   & 0.88 & 0.87 & 0.71\\
    & univ & 0.90 & 0.94 & 0.79\\
Median $z^2$ of $\alpha_{1,1}$
    & mv   & 0.52 & 0.77 & 0.96\\
    & univ & 0.49 & 0.42 & 0.87\\
$2\sigma$ coverage of $\rho_{12}$
    & mv   & 0.91 & 0.79 & 0.76\\
Median $z^2$ of $\rho_{12}$
    & mv   & 0.51 & 0.68 & 0.97\\
$2\sigma$ coverage of $b_{1,1}$
    & mv   & ---  & ---  & 0.82\\
    & univ & ---  & ---  & 0.88\\
Median $z^2$ of $b_{1,1}$
    & mv   & ---  & ---  & 0.92\\
    & univ & ---  & ---  & 0.41\\
Positive-definite curvature
    & mv   & 0.85 & 0.92 & 0.79\\
    & univ & 1.00 & 1.00 & 1.00\\
\hline
\end{tabular}
\end{table}

For MCARMA$(1,0)$, the curvature-based intervals are reasonably well
calibrated, although coverage remains somewhat below the nominal
0.95 benchmark. The multivariate $2\sigma$ coverage of
$\alpha_{1,1}$ is 0.88 and that of $\rho_{12}$ is 0.91, with median
$z^2$ values of 0.52 and 0.51, respectively. The corresponding
univariate AR coverage is 0.90 with median $z^2=0.49$. Thus, for the
simplest model, both the empirical coverage and the standardized
errors are comparatively close to the Gaussian reference values.

Calibration deteriorates as the fitted stochastic dynamics become
more complex. For MCARMA$(2,0)$, coverage of $\alpha_{1,1}$ remains
relatively high at 0.87 for the multivariate fit, but coverage of
$\rho_{12}$ falls to 0.79. For MCARMA$(2,1)$, the corresponding
coverage values are 0.71 for $\alpha_{1,1}$, 0.76 for $\rho_{12}$,
and 0.82 for $b_{1,1}$. Their median $z^2$ values are 0.96, 0.97,
and 0.92, respectively, all substantially larger than the reference
value 0.455. The local Gaussian approximation therefore understates
the total estimation error in these more weakly identified
directions.

The standardized-error diagnostic also helps distinguish poor
coverage caused mainly by the estimated standard error from poor
coverage caused by displacement of the point estimate. For example,
the median $z^2$ values for $\alpha_{1,1}$ increase from 0.52 to
0.77 and 0.96 as the multivariate model order increases. This
progression parallels the increasing complexity of the likelihood
geometry and is consistent with the identification difficulties
observed in the parameter- and model-recovery experiments. The
calibration problem should therefore not be interpreted solely as a
failure of Hessian inversion: both local curvature and displacement
of the fitted solution from the generative parameter can contribute.

The positive-definiteness of the observed information provides a
related but distinct diagnostic. The multivariate curvature is
positive definite for approximately 0.85, 0.92, and 0.79 of the
simulated MCARMA$(1,0)$, MCARMA$(2,0)$, and MCARMA$(2,1)$ fits,
respectively. Thus, even in this comparatively information-rich
simulation design, some fitted solutions do not exhibit the regular
local quadratic geometry required for conventional Hessian-based
inference. By contrast, the corresponding univariate curvature is
positive definite for essentially all fits in this experiment.

A positive-definite observed information matrix should therefore be
viewed as a useful local diagnostic rather than as evidence by
itself that the Gaussian uncertainty approximation is well
calibrated. Positive definiteness establishes that the fitted
likelihood has the expected local curvature at the reported point,
but it does not establish symmetry of the likelihood, absence of
bias, or adequate identification over the larger parameter region
relevant to finite-sample uncertainty. Conversely, failure of
positive definiteness is informative because it indicates that the
fitted likelihood does not locally behave like a regular interior
maximum in all parameter directions. This behavior is consistent
with the weak-identification mechanisms described in
Section~\ref{sec4:identifiability}.

The comparison with separate univariate fits further shows that
improved point estimation under joint multivariate modeling does not
automatically imply better nominal coverage for every individual
parameter. For $\alpha_{1,1}$, the univariate coverage rates
0.90, 0.94, and 0.79 equal or exceed the corresponding multivariate
rates 0.88, 0.87, and 0.71. This result does not contradict the
improvement in parameter and spectral point estimation demonstrated
in Section~\ref{sec:simulation.multivariate}. Joint fitting changes
the geometry and dimension of the likelihood and additionally
estimates cross-band dependence, so a more accurate point estimate
of the stochastic process need not yield a more accurate local
Gaussian approximation for every marginal parameter.

We therefore treat Hessian-based standard errors as computationally
convenient local uncertainty summaries whose reliability depends on
the dynamical regime and strength of identification, rather than as
uniformly calibrated finite-sample confidence intervals. In
well-identified regions they can provide useful uncertainty
summaries, whereas higher-order, boundary-adjacent, or otherwise
weakly identified fits may require profile-likelihood or
sampling-based methods for more complete uncertainty
characterization.

\subsection{Sensitivity to regularization, likelihood refinement,
and model selection}
\label{sec:supp.simulation.sensitivity}

The first-stage penalties and covariance loading are introduced only
to stabilize numerical optimization. To assess whether they
materially alter the resulting inference, we compare the first-stage
regularized solution with the second-stage ordinary maximum
likelihood solution using the ordinary, unloaded log-likelihood at
both parameter values. This comparison separates the statistical
likelihood from the computational regularization used to initialize
its optimization.

For a first-stage solution
$\widehat{\boldsymbol{\theta}}_{\mathrm{init}}$ and final MLE
$\widehat{\boldsymbol{\theta}}$, define the ordinary-likelihood gain
\begin{equation*}
\Delta\ell
=
\ell(\widehat{\boldsymbol{\theta}})
-
\ell(\widehat{\boldsymbol{\theta}}_{\mathrm{init}}),
\end{equation*}
where both likelihoods are evaluated without the first-stage
penalties or covariance loading. We likewise evaluate model-selection
criteria using the ordinary likelihood so that differences among
candidate orders are not contaminated by differences in the number
or type of regularized root quantities.

In the parameter-and-noise simulation study, the second-stage
ordinary-likelihood refinement has very little effect when the fitted
candidate model coincides with the generative model. Across the
examined correctly specified fits, the gain in ordinary
log-likelihood is at most approximately 0.2 nats. Thus, when the
candidate dynamics are well supported by the data, the regularized
first stage already places the optimizer close to the ordinary
likelihood maximum.

The behavior is different when an unnecessarily flexible candidate
model is fitted to data generated from a simpler process. For
example, when an MCARMA$(2,0)$ candidate is fitted to
MCARMA$(1,0)$ data, median second-stage likelihood gains of
approximately 4.7, 24.2, and 5.2 nats occur in the three examined
settings. In these cases, the additional second-order dynamical
degrees of freedom are only weakly supported by the observations,
and the soft first-stage regularization can have a substantial effect
on the part of the likelihood surface reached by the preliminary
optimization.

This contrast clarifies the role of the regularization. It is nearly
inactive in the neighborhood of a well-supported correctly specified
model but can materially influence an overparameterized candidate
whose additional poles or zeros lie in weakly identified regions.
The second-stage ordinary-likelihood refinement is therefore
important for model comparison even when it changes the correctly
specified fit only slightly. Without this refinement, an information
criterion calculated from the first-stage objective would partly
reflect the order-dependent numerical regularization rather than
only the fit of the statistical model.

A large value of $\Delta\ell$ should consequently not be interpreted
as evidence that the regularization biases the final estimator. The
regularized solution is not the reported estimator. Instead, a large
gain indicates that the first stage has fulfilled its role as a
stable numerical starting point from which maximization of the
ordinary likelihood can move to a better-supported region of the
parameter space. Conversely, the small gains observed for correctly
specified models show that the first-stage penalties can stabilize
the search without materially displacing a likelihood optimum that
is already well supported by the observations.

The use of soft rather than hard regularization is important for the
same reason. The first-stage preferred ranges discourage poorly
resolved or numerically unstable configurations but do not exclude
them. If the ordinary likelihood supports a long characteristic
timescale or any other value outside a first-stage preferred range,
the second-stage optimization remains free to move there. The covariance loading plays an analogous role:
it stabilizes likelihood calculations near nearly singular
driving-covariance configurations during initialization but is
removed before the reported MLE is obtained.

We also examine sensitivity to the model-selection criterion itself.
As an alternative to AICc, we consider the generalized information
criterion (GIC) of \citet{konishi1996gic}. GIC replaces the nominal
parameter count by an estimated effective dimension
\begin{equation*}
\widehat d_{\rm eff}
=
\operatorname{tr}
\left(
\widehat J^{-1}\widehat K
\right),
\end{equation*}
where $\widehat J$ is based on the negative Hessian of the ordinary
log-likelihood and $\widehat K$ is based on outer products of score
contributions, both evaluated at the final ordinary-likelihood
estimate. Under standard correctly specified maximum-likelihood
conditions, $\widehat d_{\rm eff}$ approaches the ordinary
parameter count.

Table~\ref{tab:controlled.gic} compares GIC with AICc in the
simulation study. At the correctly specified fits, the
median estimated effective dimensions are 25.2, 31.5, and 36.8 for
MCARMA$(1,0)$, MCARMA$(2,0)$, and MCARMA$(2,1)$, respectively,
compared with nominal parameter counts of 25, 30, and 35. Thus, on
average, the estimated complexity is reasonably close to the
ordinary parameter count. The variability of the estimated
effective dimension, however, is sufficient to affect model-order
selection.

\begin{table}[b]
\centering
\caption{Sensitivity of model-order selection to AICc and the
generalized information criterion (GIC) in the simulation
study. $\widehat d_{\rm eff}$ denotes the median estimated effective
dimension at the correctly specified fit, with the nominal parameter
count in parentheses. Negative effective dimensions are counted
across all three candidate-order fits.}
\label{tab:controlled.gic}
\begin{tabular}{lc}
\hline\hline
Quantity & Controlled study\\
\hline
$\widehat d_{\rm eff}$, $(1,0)$ & 25.2 (25)\\
$\widehat d_{\rm eff}$, $(2,0)$ & 31.5 (30)\\
$\widehat d_{\rm eff}$, $(2,1)$ & 36.8 (35)\\
5th--95th percentile, $(2,1)$ & [21.3, 46.4]\\
Negative $\widehat d_{\rm eff}$ & 78/2673\\
\hline
Correct-order recovery, AICc & 0.827\\
Correct-order recovery, GIC  & 0.671\\
\hline
Generative fraction $(2,1)$ & 0.333\\
Fraction selected $(2,1)$, AICc & 0.341\\
Fraction selected $(2,1)$, GIC  & 0.439\\
\hline
AICc--GIC disagreements & 207\\
Into $(2,1)$ / out of $(2,1)$ & 122/35\\
AICc correct / GIC correct & 163/24\\
\hline
\end{tabular}
\end{table}

Despite the broadly reasonable median effective dimensions, overall
correct-order recovery decreases from 0.827 under AICc to 0.671
under GIC. GIC also selects MCARMA$(2,1)$ for 0.439 of the data
sets, compared with a generative fraction of 0.333 and an AICc
selection fraction of 0.341. Thus, in this simulation experiment,
the estimated effective-dimension adjustment tends to shift
selection toward the most flexible candidate order.

The data-set-level disagreements make this difference clearer. AICc
and GIC select different orders for 207 data sets. Of these
disagreements, 122 move into MCARMA$(2,1)$ under GIC whereas only
35 move out of MCARMA$(2,1)$. More importantly, AICc selects the
generative order in 163 of the 207 disagreements, compared with only
24 for GIC. Thus, agreement of an aggregate selected fraction with a
population fraction would not by itself establish superior
data-set-level selection.

The poorer GIC performance is consistent with additional sampling
variability introduced through estimation of the effective
dimension. Although $\widehat d_{\rm eff}$ is close to the nominal
dimension on average, it can vary substantially across individual
fits and can even become negative, which is not meaningful as an
effective parameter count. Such variability is especially
consequential when competing information criteria are close, because
a modest change in the estimated complexity penalty can change the
selected order.

These known-truth experiments therefore provide no evidence that
replacing the nominal parameter count by the estimated effective
dimension improves order selection for the settings considered
here. We consequently use AICc for the primary model-selection
results and regard the GIC analysis as a sensitivity check rather
than as independent confirmation. More generally, this comparison
illustrates the value of assessing a model-selection criterion in
known-truth simulations before interpreting its apparent flexibility
as an advantage.

Overall, the sensitivity experiments support a clear separation
among three roles: regularization stabilizes the numerical search,
the ordinary likelihood defines the reported estimator, and AICc
provides the primary model-order comparison. The first-stage
regularization can matter materially for overflexible candidates,
but the second-stage refinement prevents those computational choices
from entering the final information criterion directly.

\subsection{Computational performance}
\label{sec:supp.simulation.computation}

We conclude the simulation study by examining the
computational cost of the proposed estimation procedure. As discussed
in Section~\ref{sec3}, the state-space representation avoids the
dense covariance calculations associated with direct
Gaussian-process likelihood evaluation and permits likelihood
evaluation with cost linear in the number of unique observation
times for fixed state dimension. The total wall-clock cost of model
fitting, however, also depends on the dimension and geometry of the
parameter space, the number of optimizer iterations, the number of
starting values and restarts, and the two-stage estimation
procedure.

Table~\ref{tab:cputime} reports average wall-clock times for the parameter-and-noise simulation corpus. The reported times correspond to a single fitted model on the computing cluster and include the two-stage estimation procedure, optimization restarts, and just-in-time compilation. The multivariate data sets contain approximately 2500 scalar measurements across five bands, whereas a corresponding individual-band time series contains approximately 500 measurements.

The computational cost increases substantially with model order for
the joint MCARMA fits. Average fitting times are approximately
147 seconds for MCARMA$(1,0)$, 792 seconds for MCARMA$(2,0)$, and
867 seconds for MCARMA$(2,1)$. Thus, the first-order five-band model
requires only a few minutes, whereas the second-order fits require
approximately 13--15 minutes under the production optimization
procedure used here. The corresponding separate univariate fits
require approximately 6, 12, and 12 seconds per band.

\begin{table}[b]
\centering
\caption{Average wall-clock time in seconds in the
simulation study. The MCARMA columns report the time for one joint
five-band fit, whereas the univariate columns report the time for
one individual-band CARMA fit. ``Fit'' includes the two-stage
estimation procedure and optimization restarts, and ``SE'' denotes
the analytic-Hessian standard-error calculation.}
\label{tab:cputime}
\begin{tabular}{lrrrr}
\hline\hline
& \multicolumn{2}{c}{MCARMA (5-band)}
& \multicolumn{2}{c}{Univariate (per band)}\\
\cline{2-3}\cline{4-5}
Order & Fit & SE & Fit & SE\\
\hline
$(1,0)$ & 147 & 43 & 6  & 11\\
$(2,0)$ & 792 & 68 & 12 & 19\\
$(2,1)$ & 867 & 89 & 12 & 16\\
\hline
\end{tabular}
\end{table}

These ratios should not be interpreted as a change from linear to
cubic scaling in the number of observations. The multivariate
likelihood is still evaluated recursively through the Kalman filter.
Rather, the joint fits combine a larger data set, a larger latent
state, and a substantially higher-dimensional optimization problem.
The five-band simulated data sets contain approximately 2500 scalar
measurements, compared with approximately 500 measurements in an
individual band. Moreover, even MCARMA$(1,0)$ contains 25 free
parameters in the five-band analysis, compared with only three for
the corresponding univariate damped random walk. The second-order
models contain still more dynamic parameters and can introduce
weakly identified likelihood directions requiring additional
optimizer iterations or restarts.

The increase in wall-clock time from MCARMA$(1,0)$ to the
second-order models is therefore driven primarily by numerical
optimization rather than by a qualitative change in the scaling of
one likelihood evaluation. For fixed numbers of bands and fixed
model order, adding observation times increases the Kalman-filter
cost approximately linearly. Increasing the number of bands or model
order, by contrast, enlarges both the state-space matrices and the
parameter space and can make optimization substantially more
demanding.

The difference between a multivariate fit and one single-band fit also reflects the additional inferential task performed by the multivariate model. A single-band analysis does not estimate any cross-band dependence and does not exploit information from the other four correlated time series. As shown in Section~\ref{sec:simulation.multivariate}, the additional computation is accompanied by improved parameter recovery in most simulation settings and substantially more consistent improvement in spectral recovery.

Likelihood-based uncertainty quantification is considerably cheaper
than obtaining the corresponding point estimate. The analytic-Hessian
standard-error calculation adds approximately 43, 68, and 89 seconds
for the three multivariate orders, compared with 11, 19, and
16 seconds for the corresponding per-band univariate fits. These
costs are smaller than the fitting times because the curvature is
evaluated only at the retained solution rather than requiring a new
search over parameter space. Once a reliable ordinary-likelihood MLE
has been obtained, local uncertainty quantification therefore
represents a comparatively modest additional computational expense.

At the same time, the timing results show where further computational
development would be most useful. The Kalman recursion itself retains
favorable scaling with light-curve length, whereas repeated
higher-dimensional optimization dominates practical runtime for the
models considered here. Improvements in initialization, parallel
multistart optimization, gradient-based numerical methods, or other
strategies for navigating the higher-order likelihood could
therefore provide substantial gains without changing the underlying
state-space likelihood.

Taken together, the simulation experiments establish several
features of the proposed framework. Model-order recovery depends
strongly on whether the relevant dynamical scales and pole--zero
structure are resolved by the observations. Joint multivariate
estimation generally improves parameter recovery and produces
larger and more consistent gains in spectral recovery. Weak
identification can arise either from limited temporal resolution or
from near pole--zero cancellation, and these mechanisms affect
model selection, parameter estimation, and uncertainty
quantification in different ways. Hessian-based uncertainty
summaries are useful in well-identified settings but are not
uniformly calibrated. The two-stage estimation procedure separates
numerical stabilization from ordinary maximum likelihood inference,
while the GIC sensitivity experiment supports the use of AICc for
the primary order comparisons in the settings considered here.

The additional computational cost of joint modeling arises largely
from the larger and more difficult optimization problem rather than
from a loss of the favorable state-space scaling with observation
length. For the five-band models considered here, complete fits
remain practical on the scale of minutes per data set. In the next
section, we illustrate the methodology using three representative
SDSS quasar data sets.

\section{Supplementary Results for the SDSS Quasar Case Studies}
\label{sec:supp.sdss}

This section provides detailed numerical results and diagnostics for
the three SDSS quasar case studies summarized in Section~\ref{sec:sdss}. The three objects were selected from the 134 eligible
Stripe~82 quasars to illustrate MCARMA$(1,0)$, MCARMA$(2,0)$, and
MCARMA$(2,1)$ fits, respectively; they are not intended to represent
the population frequencies of the three orders. We report the
complete fitted-parameter summaries, local curvature diagnostics,
detailed timescale calculations, and additional interpretation
underlying the more compact presentation in the main text.

The three parameter tables use common notation. Autoregressive and
moving-average coefficients are in inverse days, and $L$ denotes the
Cholesky factor of the driving covariance with its diagonal
parameterized on the log scale. The level offsets $\mu$ apply to the
mean-centered series and are not band magnitudes. A dagger marks a
coordinate that keeps less than 0.01 of itself in the curvature
subspace retained by the pseudoinverse; in these tables such
coordinates are driving log-variances that have collapsed toward
zero, where the likelihood is locally flat and no standard error is
reported. An asterisk marks a coordinate whose implied timescale
falls outside the interval constrained by the observing design:
2.1--667 days for Quasar~1, 1.4--446 days for Quasar~2, and 1.4--666 days
for Quasar~3.

\subsection{Quasar~1: MCARMA$(1,0)$}
\label{sec:supp.sdss.agn1}

Quasar~1 is catalog object 1995955 at right ascension $37.604^\circ$ and declination $-0.006^\circ$. Its 480 measurements span 3337 days, with a median within-band spacing of 3.0 days. The fitted stationary standard deviations are 0.174, 0.175, 0.153, 0.109, and 0.068 mag in $u$ through $z$. Relative to the median reported measurement uncertainties, these correspond to stationary signal-to-noise ratios of 2.6, 7.0, 4.6, 3.2, and 0.6. Thus the $z$ band contains less fitted intrinsic variation than measurement noise and is the least informative of the five.

The AICc value is $-1185.24$ for MCARMA$(1,0)$, $-1179.69$ for
MCARMA$(2,0)$, and $-1176.11$ for MCARMA$(2,1)$. The first-order model is preferred by 5.55 over MCARMA$(2,0)$ and 9.13 over MCARMA$(2,1)$. The modest separation is best interpreted as a lack of evidence requiring higher-order dynamics rather than strong evidence excluding them.

Table~\ref{tab:agn1.params} reports the complete parameter estimates and local standard errors. Four of the five log driving-variance coordinates do not have usable curvature-based standard errors. The leading component of the fitted driving correlation accounts for 99.0\% of its variation, leaving little curvature in directions orthogonal to that dominant component; 23 of the 25 fitted coordinates have positive local curvature. This behavior motivates the caution concerning near-singular driving covariance matrices discussed in Section~\ref{sec:sdss.coherence}.

Under MCARMA$(1,0)$ each band has one real autoregressive root. The corresponding decay timescales are 374, 380, 432, 493, and 408 days in $u$ through $z$, all within the approximate 2.1--667 day interval informed by the observing design. Although the point estimates tend to increase from the blue bands through $i$, the autoregressive coefficients are determined only to roughly two standard errors, so the ordering should not be interpreted as established for this individual object.

\begin{table}[b]
\centering
{\footnotesize
\caption{Parameter estimates and standard errors for Quasar~1 under the selected MCARMA$(1,0)$ model, read down the left block and then the right. $a_{j,k}$ is the $k$th autoregressive coefficient in band $j$. $L_{jj}$ is the log driving variance in band $j$ and $L_{jl}$ the Cholesky off-diagonal coupling bands $j$ and $l$. $\mu_{j}$ is the level offset of the mean-centered band. An asterisk marks a coordinate whose value falls outside the timescale range the observing design constrains. A dagger marks a driving log-variance that has collapsed toward zero. The likelihood is flat in that direction, so the pseudoinverse drops it and no standard error is reported.}
\label{tab:agn1.params}
\begin{tabular}{lrrc@{\hspace{1.5em}}lrrc}
\hline\hline
Parameter & Estimate & SE & & Parameter & Estimate & SE & \\
\hline
$a_{u,1}$ & 0.002672 & 0.001278 &  & $L_{ir}$ & 0.0005736 & 0.0003115 &  \\
$a_{g,1}$ & 0.002635 & 0.001234 &  & $L_{ii}$ & -41.69 & \ldots & $\dagger\ast$ \\
$a_{r,1}$ & 0.002314 & 0.001189 &  & $L_{zu}$ & 0.004735 & 0.00146 &  \\
$a_{i,1}$ & 0.002028 & 0.001072 &  & $L_{zg}$ & -0.0001103 & 0.001989 &  \\
$a_{z,1}$ & 0.002451 & 0.001673 &  & $L_{zr}$ & -0.0006783 & 0.000869 &  \\
$L_{uu}$ & -8.734 & 0.284 & $\ast$ & $L_{zi}$ & 3.25e-10 & 0.0008616 &  \\
$L_{gu}$ & 0.01268 & 0.001694 &  & $L_{zz}$ & -80.86 & \ldots & $\dagger\ast$ \\
$L_{gg}$ & -50.63 & \ldots & $\dagger\ast$ & $\mu_{u}$ & -0.01872 & 0.07689 &  \\
$L_{ru}$ & 0.01028 & 0.001395 &  & $\mu_{g}$ & -0.01195 & 0.07751 &  \\
$L_{rg}$ & 0.0002695 & 0.004406 &  & $\mu_{r}$ & -0.01289 & 0.07201 &  \\
$L_{rr}$ & -12.8 & \ldots & $\dagger\ast$ & $\mu_{i}$ & -0.015 & 0.05447 &  \\
$L_{iu}$ & 0.00692 & 0.0009461 &  & $\mu_{z}$ & -0.03186 & 0.03362 &  \\
$L_{ig}$ & 9.327e-05 & 0.001623 &  &  &  &  &  \\
\hline
\end{tabular}}
\end{table}

\subsection{Quasar~2: MCARMA$(2,0)$}
\label{sec:supp.sdss.agn2}

Quasar~2 is catalog object 1465040 at right ascension $334.522^\circ$ and declination $-0.634^\circ$. Its 467 measurements span 2230 days with a median within-band spacing of 2.0 days. The fitted stationary standard deviations are 0.174, 0.223, 0.175, 0.142, and 0.138 mag, giving stationary signal-to-noise ratios of 2.0, 8.0, 5.2, 3.9, and 1.2 in $u$ through $z$.

The AICc values are $-1087.16$ for MCARMA$(1,0)$, $-1095.46$ for MCARMA$(2,0)$, and $-1084.79$ for MCARMA$(2,1)$. Thus MCARMA$(2,0)$ is preferred by 8.30 over the first-order model and by 10.67 over MCARMA$(2,1)$. The selected middle order is informative: the data support an additional autoregressive pole but do not support the moving-average component.

The complete parameter estimates are given in Table~\ref{tab:agn2.params}. As for Quasar~1, four of the five log driving-variance coordinates lack usable curvature-based standard errors, and the leading component of the fitted driving correlation accounts for 99.2\% of its variation. Quasar~2 is the least well conditioned of the three examples, with positive local curvature for 18 of 30 fitted coordinates.

Each band has two real autoregressive roots. The slow timescales are 295, 384, 471, 525, and 168 days, and the fast timescales are 23.6, 24.5, 19.0, 7.4, and 168 days in $u$ through $z$. The 7--25 day fast scales in the first four bands are several times the two-day median spacing and well below the baseline, so they represent an observationally interpretable second dynamical scale. In contrast, the slow $r$- and $i$-band timescales exceed the approximate 446-day upper bound implied by the baseline and should be regarded as extrapolations beyond the most informative temporal range. In $z$ the two roots coincide at 168 days, placing the fit at the critically damped boundary between distinct-real-root and complex-root behavior.

\begin{table}[b]
\centering
{\footnotesize
\caption{Parameter estimates and standard errors for Quasar~2 under the selected MCARMA$(2,0)$ model, read down the left block and then the right. $a_{j,k}$ is the $k$th autoregressive coefficient in band $j$. $L_{jj}$ is the log driving variance in band $j$ and $L_{jl}$ the Cholesky off-diagonal coupling bands $j$ and $l$. $\mu_{j}$ is the level offset of the mean-centered band. An asterisk marks a coordinate whose value falls outside the timescale range the observing design constrains. A dagger marks a driving log-variance that has collapsed toward zero. The likelihood is flat in that direction, so the pseudoinverse drops it and no standard error is reported.}
\label{tab:agn2.params}
\begin{tabular}{lrrc@{\hspace{1.5em}}lrrc}
\hline\hline
Parameter & Estimate & SE & & Parameter & Estimate & SE & \\
\hline
$a_{u,1}$ & 0.04578 & 0.03114 &  & $L_{rr}$ & -41.29 & \ldots & $\dagger\ast$ \\
$a_{u,2}$ & 0.0001437 & 0.0001005 &  & $L_{iu}$ & 0.001161 & 0.001067 &  \\
$a_{g,1}$ & 0.04337 & 0.0185 &  & $L_{ig}$ & 0.0002328 & 0.0002558 &  \\
$a_{g,2}$ & 0.0001062 & 5.812e-05 &  & $L_{ir}$ & 2.236e-09 & 0.0002573 &  \\
$a_{r,1}$ & 0.05467 & 0.0245 & $\ast$ & $L_{ii}$ & -69.41 & \ldots & $\dagger\ast$ \\
$a_{r,2}$ & 0.0001116 & 7.347e-05 & $\ast$ & $L_{zu}$ & 0.0001329 & 8.878e-05 &  \\
$a_{i,1}$ & 0.1362 & 0.1282 & $\ast$ & $L_{zg}$ & 1.981e-06 & 1.696e-05 &  \\
$a_{i,2}$ & 0.0002555 & 0.0002676 & $\ast$ & $L_{zr}$ & -1.768e-11 & 1.408e-05 &  \\
$a_{z,1}$ & 0.01194 & 0.008221 &  & $L_{zi}$ & 2.886e-14 & 1.391e-05 &  \\
$a_{z,2}$ & 3.888e-05 & 2.379e-05 &  & $L_{zz}$ & -122 & \ldots & $\dagger\ast$ \\
$L_{uu}$ & -14.74 & 1.34 & $\ast$ & $\mu_{u}$ & -0.03695 & 0.08723 &  \\
$L_{gu}$ & 0.0006753 & 0.0002692 &  & $\mu_{g}$ & -0.02452 & 0.1199 &  \\
$L_{gg}$ & -22.14 & \ldots & $\dagger\ast$ & $\mu_{r}$ & -0.03637 & 0.1003 &  \\
$L_{ru}$ & 0.0006025 & 0.0002592 &  & $\mu_{i}$ & -0.04264 & 0.08294 &  \\
$L_{rg}$ & 0.0001112 & 7.842e-05 &  & $\mu_{z}$ & -0.03623 & 0.06919 &  \\
\hline
\end{tabular}}
\end{table}

\subsection{Quasar~3: MCARMA$(2,1)$}
\label{sec:supp.sdss.agn3}

Quasar~3 is catalog object 3064008 at right ascension $49.128^\circ$ and declination $-1.051^\circ$. It has 676 measurements over 3331 days with a median within-band spacing of 2.0 days. Its fitted stationary standard deviations are 0.156, 0.134, 0.103, 0.091, and 0.073 mag, corresponding to stationary signal-to-noise ratios of 4.2, 7.9, 7.9, 6.2, and 2.7. It is therefore the best measured of the three examples; even the $z$ band has fitted intrinsic variability exceeding its typical measurement noise.

The AICc values are $-2228.36$ for MCARMA$(1,0)$, $-2193.42$ for MCARMA$(2,0)$, and $-2443.49$ for MCARMA$(2,1)$. MCARMA$(2,1)$ is preferred by 215.13 over the next-best model. MCARMA$(2,0)$ is itself 34.94 AICc units worse than MCARMA$(1,0)$, showing that the decisive improvement is associated with the moving-average component rather than simply the additional autoregressive order.

Table~\ref{tab:agn3.params} reports the complete estimates. Positive local curvature is obtained for 33 of 35 fitted coordinates. The five moving-average coefficients are among the best determined dynamic parameters in the three case studies, each separated from zero by approximately three to four local standard errors.

The fitted autoregressive roots are real in every band. The slow timescales are 433, 429, 467, 517, and 540 days, while the fast timescales are 0.003, 0.009, 0.005, 0.022, and 0.585 days. The latter are far below the two-day sampling scale and cannot be interpreted as separately resolved timescales. The moving-average zeros correspond to 2.57, 2.30, 2.61, 1.07, and 0.119 days; only the first three exceed the approximate 1.4-day lower bound associated with the cadence. The appropriate interpretation is therefore that the observations resolve a slow component near 500 days together with a modification of the short-timescale spectrum, but do not separately resolve the fast pole and the nearby moving-average structure.

\begin{table}[b]
\centering
{\footnotesize
\caption{Parameter estimates and standard errors for Quasar~3 under the selected MCARMA$(2,1)$ model, read down the left block and then the right. $a_{j,k}$ is the $k$th autoregressive coefficient in band $j$. $b_{j,k}$ is the $k$th moving-average coefficient in that band. $L_{jj}$ is the log driving variance in band $j$ and $L_{jl}$ the Cholesky off-diagonal coupling bands $j$ and $l$. $\mu_{j}$ is the level offset of the mean-centered band. An asterisk marks a coordinate whose value falls outside the timescale range the observing design constrains. A dagger marks a driving log-variance that has collapsed toward zero. The likelihood is flat in that direction, so the pseudoinverse drops it and no standard error is reported.}
\label{tab:agn3.params}
\begin{tabular}{lrrc@{\hspace{1.5em}}lrrc}
\hline\hline
Parameter & Estimate & SE & & Parameter & Estimate & SE & \\
\hline
$a_{u,1}$ & 316.6 & 108.9 &  & $L_{ru}$ & 1.381 & 0.6637 &  \\
$a_{u,2}$ & 0.7306 & 0.4303 &  & $L_{rg}$ & -0.09275 & 0.07288 &  \\
$a_{g,1}$ & 114.9 & 34.3 &  & $L_{rr}$ & -29.98 & \ldots & $\dagger\ast$ \\
$a_{g,2}$ & 0.2676 & 0.1461 &  & $L_{iu}$ & 0.2448 & 0.1102 &  \\
$a_{r,1}$ & 213.3 & 95.15 &  & $L_{ig}$ & -0.04104 & 0.02976 &  \\
$a_{r,2}$ & 0.4568 & 0.2977 &  & $L_{ir}$ & 1.762e-07 & 0.01358 &  \\
$a_{i,1}$ & 45.81 & 19.1 &  & $L_{ii}$ & -2327 & \ldots & $\dagger\ast$ \\
$a_{i,2}$ & 0.08859 & 0.05869 &  & $L_{zu}$ & 0.006723 & 0.004031 &  \\
$a_{z,1}$ & 1.711 & 0.9926 &  & $L_{zg}$ & -0.001518 & 0.001271 &  \\
$a_{z,2}$ & 0.003162 & 0.002358 &  & $L_{zr}$ & 6.27e-09 & 0.0006614 &  \\
$b_{u,1}$ & 0.3886 & 0.1149 &  & $L_{zi}$ & -5.635e-10 & 0.000499 &  \\
$b_{g,1}$ & 0.4347 & 0.1194 &  & $L_{zz}$ & -7.525e+05 & \ldots & $\dagger\ast$ \\
$b_{r,1}$ & 0.3837 & 0.1218 &  & $\mu_{u}$ & -0.09324 & 0.0694 &  \\
$b_{i,1}$ & 0.937 & 0.2835 &  & $\mu_{g}$ & -0.07245 & 0.06068 &  \\
$b_{z,1}$ & 8.428 & 3.057 &  & $\mu_{r}$ & -0.05278 & 0.04774 &  \\
$L_{uu}$ & 2.316 & 0.7843 &  & $\mu_{i}$ & -0.04307 & 0.0439 &  \\
$L_{gu}$ & 1.024 & 0.3541 &  & $\mu_{z}$ & -0.03169 & 0.03427 &  \\
$L_{gg}$ & -873.7 & \ldots & $\dagger\ast$ &  &  &  &  \\
\hline
\end{tabular}}
\end{table}

\subsection{Additional spectral comparisons}
\label{sec:supp.sdss.psd}

The main text compares the marginal PSDs obtained from the selected joint five-band MCARMA fit with those obtained by fitting the corresponding univariate CARMA model to each band separately. These comparisons should not be interpreted as a truth-based assessment, because the generative spectra are unknown for the SDSS data. Rather, they show how cross-band information changes marginal spectral inference.

For comparison with the joint MCARMA analysis, each band is also fitted independently, with its univariate CARMA order selected separately by AICc from CARMA$(1,0)$, CARMA$(2,0)$, and CARMA$(2,1)$. Thus, the single-band PSDs are not conditioned on the order selected by the joint model. For Quasar~1, the joint analysis selects MCARMA$(1,0)$ and therefore yields marginal DRW spectra, whereas several independently selected single-band fits exhibit higher-order spectral structure, most prominently in the weak $z$ band. For Quasar~2, the independently selected fits likewise produce substantially more heterogeneous spectral shapes across filters than the joint MCARMA$(2,0)$ fit. Quasar~3 provides a contrasting case in which several individual-band spectra closely track the joint estimates. Because the underlying stochastic processes are unknown for these real data, these comparisons do not establish which representation is closer to the truth; rather, they illustrate how sharing information across dependent bands can materially affect model-order selection and marginal spectral inference.

For MCARMA$(2,1)$, the flattening visible over the plotted frequency range is a finite-frequency consequence of the moving-average zero and should not be confused with the asymptotic spectral slope. Since $p-q=1$, the ultimate high-frequency behavior of MCARMA$(2,1)$ is proportional to $f^{-2}$ even when the local slope over the observationally relevant frequency range is substantially shallower.

\subsection{Driving covariance and coherence diagnostics}
\label{sec:supp.sdss.coherence}

Under the diagonal dynamic structure used in the paper, magnitude-squared coherence is frequency-independent and equals the squared correlation between the corresponding stochastic drivers. Across the three objects the estimated coherences range from 0.909 to 1.000. Quasar~3 shows the clearest decrease with increasing wavelength separation; Quasar~2 does not show a monotone ordering.

The fitted driving covariance is nearly rank one in all three objects: its leading component accounts for 99.0--99.2\% of the driving variation. Consequently, estimates near the boundary of the correlation space require caution. In particular, an estimated coherence of 1.000 with a near-zero delta-method standard error reflects the boundary constraint and near-singular covariance geometry rather than essentially perfect finite-sample precision. This same geometry explains why several driving-variance coordinates have weak or unusable local curvature even when the fitted marginal spectra are well behaved.

\subsection{Summary of local identifiability diagnostics}
\label{sec:supp.sdss.identifiability}

The three examples illustrate different relationships between model selection and parameter identifiability. For Quasar~1, all fitted decay timescales lie within the temporal range informed by the data, although the AICc separation from the higher-order models is modest. For Quasar~2, AICc supports a second autoregressive pole, but two slow-timescale estimates extend beyond the range strongly constrained by the finite baseline. For Quasar~3, the moving-average model is selected decisively even though the fitted fast autoregressive timescales lie far below the sampling scale. Thus selection of a higher-order stochastic structure does not imply that every individual root or zero is separately resolved. In such cases the fitted PSD can provide a more stable and scientifically interpretable summary of the jointly estimated dynamics than the raw parameter coordinates themselves.

\bibliographystyle{spr-ims-nameyear}
\bibliography{reference}

\end{document}